\documentclass[%
 reprint,
 amsmath,amssymb,
 aip,
]{revtex4-1}

\usepackage{graphicx}
\usepackage{dcolumn}
\usepackage{bm}
\usepackage{color}

\begin{document}

\preprint{APS/123-QED}

\title{Interface Dark Excitons at Sharp Lateral Two-Dimensional Heterostructures}

\author{Hamidreza Simchi}
\email{simchi@alumni.iust.ac.ir}
\affiliation {Department of Physics, Iran University of Science and Technology, Narmak, Tehran 16844, Iran} \affiliation{Semiconductor Technoloy Center, P.O.Box 19575-199, Tehran, Iran}

\date{\today}

\begin{abstract}
We study the dark excitons at the interface of a sharp lateral heterostructure of two-dimensional transition metal dichalcogenides (TMDs).  By introducing a low-energy effective Hamiltonian model, we find the energy dispersion relation of exciton and show how it depends on the onsite energy of composed materials and their spin-orbit coupling strengths. It is shown that the effect of the geometrical structure of the interface, as a deformation gauge field (pseudo-spin-orbit coupling), should be considered in calculating the binding energy of exciton. By discretization of the real-space version of the dispersion relation on a triangular lattice, we show that the binding energy of exciton depends on its distance from the interface line. For exciton near the interface, the binding energy is equal to 0.36 eV, while for the exciton far enough from the interface, it is equal to 0.26 eV. Also, it has been shown that for a zigzag interface the binding energy increases by 0.34 meV compared to an armchair interface due to the pseudo-spin-orbit interaction (gauge filed). The results can be used for designing 2D-dimensional-lateral-heterostructure- based optoelectronic devices to improve their characteristics.
\end{abstract}

\pacs{73.63.-b, 71.35 Cc, 71.70.Ej}
\keywords{2D materials, Exciton, Binding energy, pseudo-spin orbit coupling }
\maketitle


\section{Introduction}
Direct band gap semiconductors are one of the main candidates of photon absorption or emission. If the energy of photons be equal or greater than the energy band gap of the semiconductors they will excite the electrons from valence band to the conduction band. Also, the photons will be emitted by coming back the excited electrons to the valence band. By adding the n (p)-type impurities to semiconductors new energy levels are created near the conduction (valence) band edge inside the band gap of semiconductors. The electrons and holes can be exited to the new energy levels in photon absorption process or coming back to the valence band in photon emission process. Another mechanism for creating an energy level inside the band gap is the creation of exciton.  An exciton can be created by coupling an electron with a hole. If the spin of electron and hole be not (be) same a dark (bright)  ecxiton is created. Also, by absorption of photons near the junction area of a reversed bias pn-junction the created electrons (holes) will be injected to the n (p)-region and in consequence a photocurrent is created. If the band gap of the semiconductor be small, such as Indium Antimonite (InSb), the device should be cooled to very low temperature. Therefore, the direct or indirect type of band gap, the value of the band gap, the working temperature of device and the creation of energy levels in the band gap by adding the impurities or creation of exciton are important factors in designing the optoelectronic devices.\\

A homojunction can be created by doping an impurity in a semiconductor material with specific band gap energy. The interface of two dissimilar semiconductors forms a heterojunction. The combination of multiple heterojunctions together in a device is called heterostructure. The semiconductors band alignment in heterojunctions can be categorized into three different types. In type I (straddling gap), the conduction band minimum (CBM) of one material is contained (nested) inside the band gap of other material. If the band-gap of one material is rested inside the band gap of the other material, the type II alignment (staggered gap) is formed. In type III (broken gap), the CBM of one material is equal to the valence band maximum (VBM) of the other material. Therefore, the different types of heterostructures can be created by using the heterojunctions  and the band gap engineering technique. The new heterostructures can be considered as new candidate for designing the optoelectronic devices.\\
\noindent     Recently, two dimensional materials (2DMs) have been considered as the new candidate for manufacturing heterostructurs. Different two-dimensional materials with honeycomb structures have been investigated. Hexagonal boron nitride (h-BN) \cite{R1,R2}, transition metal dichalcogenides (TMDs) \cite{R3,R4}, black phosphorus (BP) \cite{R5,R6} and silicene \cite{R7} are the most widely studied 2DMs. By vertical (vertical heterostructures (VHSs)) or lateral (lateral heterostructures (LHSs)) integration of 2DMs, an artificial heterostructured 2D layer can be fabricated. The single-, double-, and multi-step chemical vapor deposition (CVD) approaches have been used to grow large -area, sharp 2D heterostructures \cite{R8,R9,Rten,R11}. Also, using the techniques, researchers could grow hetero-triangles composed of a central TMD and an outer triangular ring of another TMD \cite{R8, R12, R13}. The other structures such as truncated triangles, hexagons, hexagrams \cite{R14} and more complex patterned structures have also been reported \cite{R15}.

\noindent     The built-in potential in pn-junction and its increasing rate under the reversed bias condition are important parameters in designing the homojundtion-based optoelectronic devices. Similarly, the band alignment (or band offset) across the junction is an important parameter in material design. The band offset in VHSs \cite{R16,R17} and in LHSs \cite{R18} have been studied by using the density functional theory (DFT). Ozcelik et al., have introduced the periodic table of heterostructures (HSs) based on the band offset between them \cite{R19}. For studying the properties of HSs the tight-binding approach has been used for not only commensurate HSs \cite{R20,R21} but also for incommensurate types \cite{R22}.   

\noindent    Different types of devices can be designed and manufactured after acquisition of design and manufacturing technologies of two dimensional Heterostures (2DHSs). For example, by using monolayer WSe${}_{2}$-WS${}_{2}$ HS, high mobility field-effect transistors (FETs), complementary metal-oxide semiconductor (CMOS) and superior photovoltaic devices have been demonstrated \cite{R23}. A light-emitting device (LED) with large conversion efficiency has been built by using the lateral WSe${}_{2}$-MoS${}_{2}$ HS \cite{R24}. It has been shown that one can manufacture a photodiode with high photodetectivity of ${10}^{13}$ Jones and short response time of sub-$100\ \mu s$ by using the LHS of graphene and thin amorphous carbon \cite{R25}. Amani et al., have reported near-unity photon quantum yield in TMD monolayer, which led to the development of highly efficient optoelectronic devices \cite{R26}. It has been shown that Mo- and W-based TMDs have bright and dark exciton ground states, respectively, due to the reversal of spins in the conduction band \cite{R27,R28}. 

\noindent       As it has been mentioned above, exciton energy level is placed inside the energy band gap region. Therefore, the creation of exciton in 2DHSs is important from the 2DHS-based optoelectronic devices point of view. The long-lived interlayer excitons in monolayer MoSe${}_{2}$-WSe${}_{2}$ HS have been reported \cite{R29}, and the light-induced exciton spin Hall effect in van der Waals HSs has been studied \cite{R30}. Latini et al., have studied the role of the dielectric screening on the properties of excitons in van der Waals HSs \cite{R31}. Lau et al., have studied the interface excitons at lateral heterojunctions in monolayer semiconductors and shown that the competition between the lattice and Coulomb potentials implies the properties of exciton at the interface \cite{R33}.\\

One of the important properties of 2DMs is the effect of its boundary on the electronic properties of the nanoribbons composed by the materials. Therefore, it is expected that by changing the electronic property of the nanoribbons, due to the change in their boundaries, the photon absorption and emission by the nanoribbons change, too. It means that in addition to the previous important parameters, the boundary condition or the geometry of two-dimensional nanoribbons  be another important parameter in designing and manufacturing the 2DHS-basde optoelectronic devices. For example, it has been shown that the plane integrated modular wave functions of the VBM and CBM for different LHSs with long armchair and zigzag interfaces are localized on Mo-side and W-side, respectively \cite{R32}. Also, it has been shown that the wave function behavior is quite different in LHSs with zigzag interface compared to armchair interface\cite{R32}.For studying the effect of geometry on the wave function, the normal derivative is substituted by the covariant derivative. It is done by introducing a deformation gauge filed $\vec{A}$ which is subtracted from the momentum operate in the Hamiltonian of the system\cite{R32}.\\

The above short review shows that band gap energy, band offset voltage, type of alignment, interface structure, and the type of  chalcogenide in the lateral M${}_{i}$Xj-M${}_{k}$X${}_{l}$ (M=transition metal, X=chalcogenide) heterostructures are important in designing an exciton-based optoelectronic devices. The importance of these factors motivated us to study the relationship between the energy of exciton, its binding energy, and the mentioned parameters in the LHSs. 

\noindent      In this paper, we consider monolayer LHSs of transition metal dichalcogenide with armchair and zigzag interfaces. By introducing a low-energy effective Hamiltonian model,we find the energy dispersion relation of dark exciton and show how it depends on the onsite energy of composed materials and their spin-orbit coupling strengths. Using the real-space version of exciton dispersion energy relation and its discretizing on a triangular lattice we find the binding energy of exciton. It is shown that the binding energy depends on the distance of the exciton from the interface line which is governed by the competition between lattice and Coulomb potentials. We show that the effect of the geometrical structure of the interface appears as a deformation gauge field and increases the binding energy of exciton in the zigzag interface. The structure of the article is as follows: Section II includes the Hamiltonian model. The numerical calculations are provided in section III. The results and discussion and summary are provided in sections IV and V, respectively.

\section{Hamiltonian model}

\noindent 

\noindent       Let us, consider a lateral heterostructure of MoX${}_{2}$-WX${}_{2}$ with an armchair interface. Since the plane integrated modular wave function of VBM and CBM are localized on W-side and Mo-side, respectively \cite{R32}, near $K\ (K^{\prime})$-point we can consider the low-energy two-band Hamiltonian model (Appendix A) in both sides for studying the behavior of electrons and holes. If the wave functions of Mo (W)-side are called ${\phi }^{M(W)}_1$ and ${\phi }^{M(W)}_2$ , a new base function $\psi ={({\phi }^M_1{\phi }^W_1,\ {\phi }^M_1{\phi }^W_2{,\ \phi }^M_2{\phi }^W_1,\ {\ \phi }^M_2{\phi }^W_1)}^T$ can be defined such that 

\noindent 
\begin{widetext}
\begin{equation} \label{GrindEQ__1_} 
\left( \begin{array}{ccc}
H^{Mo}_{11}-H^W_{11} & -H^W_{12} &  \begin{array}{cc}
H^{Mo}_{12} & 0 \end{array}
 \\ 
-H^W_{21} & H^{Mo}_{11}-H^W_{22} &  \begin{array}{cc}
0 & H^{Mo}_{12} \end{array}
 \\ 
 \begin{array}{c}
H^{Mo}_{21} \\ 
0 \end{array}
 &  \begin{array}{c}
0 \\ 
H^{Mo}_{21} \end{array}
 &  \begin{array}{c}
 \begin{array}{cc}
H^{Mo}_{22}-H^W_{11} & -H^W_{12} \end{array}
 \\ 
 \begin{array}{cc}
-H^W_{21} & H^{Mo}_{22}-H^W_{22} \end{array}
 \end{array}
 \end{array}
\right)\left( \begin{array}{c}
{\phi }^M_1{\phi }^W_1 \\ 
{\phi }^M_1{\phi }^W_2 \\ 
 \begin{array}{c}
{\phi }^M_2{\phi }^W_1 \\ 
{\ \phi }^M_2{\phi }^W_1 \end{array}
 \end{array}
\right)=\left(E^{Mo}-E^W\right)\left( \begin{array}{c}
{\phi }^M_1{\phi }^W_1 \\ 
{\phi }^M_1{\phi }^W_2 \\ 
 \begin{array}{c}
{\phi }^M_2{\phi }^W_1 \\ 
{\ \phi }^M_2{\phi }^W_1 \end{array}
 \end{array}
\right) 
\end{equation} 
\end{widetext}

\noindent where, $H^{Mo}_{e,\uparrow }=\left( \begin{array}{cc}
H^{Mo}_{11} & H^{Mo}_{12} \\ 
H^{Mo}_{21} & H^{Mo}_{22} \end{array}
\right)$ and $H^W_{e,\uparrow }=\left( \begin{array}{cc}
H^W_{11} & H^W_{12} \\ 
H^W_{21} & H^W_{22} \end{array}
\right)$. It should be noted that the Hamiltonian of spin-down hole in W-side is given by $H^W_{h,\downarrow }=-H^W_{e,\uparrow }$ . Since, the spin of electron is up and the spin of the hole is down th etotal spin of the exciton is zero. It means that the exciton is dark. Now, we introduce four eigenfunctions as below
\begin{widetext}
\begin{equation} \label{GrindEQ__2_} 
{\psi }_1={\left(cos{(\theta }_1/2)\mathrm{cos}\mathrm{}({\theta }_2/2),\ cos{(\theta }_1/2)e^{i{\varphi }_2}\mathrm{sin}\mathrm{}({\theta }_2/2),\ e^{i{\varphi }_1}\mathrm{sin}\mathrm{}({\theta }_1/2)\mathrm{cos}\mathrm{}({\theta }_2/2),e^{i{\varphi }_1}\mathrm{sin}\mathrm{}({\theta }_1/2)e^{i{\varphi }_2}\mathrm{sin}\mathrm{}({\theta }_2/2)\right)}^T 
\end{equation} 
\end{widetext}
\begin{widetext}
\begin{equation} \label{GrindEQ__3_} 
{\psi }_2={\left(e^{-i{\varphi }_1}{\mathrm{sin} \left(\frac{{\theta }_1}{2}\right)\ }e^{-i{\varphi }_2}{\mathrm{sin} \left(\frac{{\theta }_2}{2}\right)\ },{-e}^{-i{\varphi }_1}{\mathrm{sin} \left(\frac{{\theta }_1}{2}\right)\ }{\mathrm{cos} \left(\frac{{\theta }_2}{2}\right)\ },\ -cos{(\theta }_1/2)e^{-i{\varphi }_2}{\mathrm{sin} \left(\frac{{\theta }_2}{2}\right)\ },cos{(\theta }_1/2){\mathrm{cos} \left(\frac{{\theta }_2}{2}\right)\ }\right)}^T\  
\end{equation}
\end{widetext} 
\begin{widetext}
\begin{equation} \label{GrindEQ__4_} 
{\psi }_3={\left(cos{(\theta }_1/2)e^{-i{\varphi }_2}\mathrm{sin}\mathrm{}({\theta }_2/2),\ -cos{(\theta }_1/2)\mathrm{cos}\mathrm{}({\theta }_2/2),\ e^{i{\varphi }_1}\mathrm{sin}\mathrm{}({\theta }_1/2)e^{-i{\varphi }_2}\mathrm{sin}\mathrm{}({\theta }_2/2),-e^{i{\varphi }_1}\mathrm{sin}\mathrm{}({\theta }_1/2)\mathrm{cos}\mathrm{}({\theta }_2/2)\right)}^T\  
\end{equation}
\end{widetext} 
\begin{widetext}
\begin{equation} \label{GrindEQ__5_} 
{\psi }_4={\left(e^{-i{\varphi }_1}{\mathrm{sin} \left(\frac{{\theta }_1}{2}\right)\ }\mathrm{cos}\mathrm{}({\theta }_2/2),\ e^{-i{\varphi }_1}{\mathrm{sin} \left(\frac{{\theta }_1}{2}\right)\ }e^{i{\varphi }_2}\mathrm{sin}\mathrm{}({\theta }_2/2),\ -cos{(\theta }_1/2)\mathrm{cos}\mathrm{}({\theta }_2/2),-cos{(\theta }_1/2)e^{i{\varphi }_2}\mathrm{sin}\mathrm{}({\theta }_2/2)\right)}^T\  
\end{equation} 
\end{widetext}
Here, ${\theta }_1$ (${\theta }_2$) and ${\varphi }_1$ (${\varphi }_2$) are attributed to Mo (W)-side and their definitions are provided in Appendix A.

\noindent 

\noindent Using the results of appendix A, the low-energy Hamiltonian of lateral heterostructure can be written as
\begin{widetext}
\begin{equation} \label{GrindEQ__6_} 
H=\left( \begin{array}{ccc}
\frac{{\Delta }_{1-}{\Delta }_2}{2} & -a_2t_2k_2e^{-i{\varphi }_2} &  \begin{array}{cc}
a_1t_1k_1e^{-i{\varphi }_1} & 0 \end{array}
 \\ 
-a_2t_2k_2e^{i{\varphi }_2} & \frac{{\Delta }_{1+}{\Delta }_2}{2}-{\lambda }_2 &  \begin{array}{cc}
0 & a_1t_1k_1e^{-i{\varphi }_1} \end{array}
 \\ 
 \begin{array}{c}
a_1t_1k_1e^{i{\varphi }_1} \\ 
0 \end{array}
 &  \begin{array}{c}
0 \\ 
a_1t_1k_1e^{i{\varphi }_1} \end{array}
 &  \begin{array}{c}
 \begin{array}{cc}
-\frac{{\Delta }_{1+}{\Delta }_2}{2}+{\lambda }_1 & -a_2t_2k_2e^{-i{\varphi }_2} \end{array}
 \\ 
 \begin{array}{cc}
-a_2t_2k_2e^{i{\varphi }_2} & -\frac{{\Delta }_{1-}{\Delta }_2}{2}+{\lambda }_1-{\lambda }_2 \end{array}
 \end{array}
 \end{array}
\right) 
\end{equation} 
\end{widetext}
where, the subscript 1 (2) is attributed to Mo (W)-side. Therefore, it can be easily shown that
\begin{equation}
H{\psi }_1=(E_1-E_2){\psi }_1 
\end{equation} 
\begin{equation}                                                                                                                               
H{\psi }_2=(-E_1+E_2+{\lambda }_1-{\lambda }_2){\psi }_2                                                                                                            
\end{equation}
\begin{equation}
H{\psi }_3=(E_1+E_2-{\lambda }_2){\psi }_3   
\end{equation}
\begin{equation}                                                                                                                     
H{\psi }_4=({-E}_1-E_2+{\lambda }_1){\psi }_4                                                                                                                    
\end{equation}
It is obvious that the Hamiltonian matrix, $H$, can be diagonalized by using the matrix $P=({\psi }_1,\ {\psi }_2,{\psi }_3,{\psi }_4)$ i.e.,

\noindent 
\begin{widetext}
\begin{equation} \label{GrindEQ__8_} 
P=\left( \begin{array}{ccc}
cos{(\theta }_1/2)\mathrm{cos}\mathrm{}({\theta }_2/2) & e^{-i{\varphi }_1}{\mathrm{sin} \left(\frac{{\theta }_1}{2}\right)\ }e^{-i{\varphi }_2}{\mathrm{sin} \left(\frac{{\theta }_2}{2}\right)\ } &  \begin{array}{cc}
cos{(\theta }_1/2)e^{-i{\varphi }_2}\mathrm{sin}\mathrm{}({\theta }_2/2) & e^{-i{\varphi }_1}{\mathrm{sin} \left(\frac{{\theta }_1}{2}\right)\ }\mathrm{cos}\mathrm{}({\theta }_2/2) \end{array}
 \\ 
cos{(\theta }_1/2)e^{i{\varphi }_2}\mathrm{sin}\mathrm{}({\theta }_2/2) & {-e}^{-i{\varphi }_1}{\mathrm{sin} \left(\frac{{\theta }_1}{2}\right)\ }{\mathrm{cos} \left(\frac{{\theta }_2}{2}\right)\ } &  \begin{array}{cc}
-cos{(\theta }_1/2)\mathrm{cos}\mathrm{}({\theta }_2/2) & e^{-i{\varphi }_1}{\mathrm{sin} \left(\frac{{\theta }_1}{2}\right)\ }e^{i{\varphi }_2}\mathrm{sin}\mathrm{}({\theta }_2/2) \end{array}
 \\ 
 \begin{array}{c}
e^{i{\varphi }_1}\mathrm{sin}\mathrm{}({\theta }_1/2)\mathrm{cos}\mathrm{}({\theta }_2/2) \\ 
e^{i{\varphi }_1}\mathrm{sin}\mathrm{}({\theta }_1/2)e^{i{\varphi }_2}\mathrm{sin}\mathrm{}({\theta }_2/2) \end{array}
 &  \begin{array}{c}
-cos{(\theta }_1/2)e^{-i{\varphi }_2}{\mathrm{sin} \left(\frac{{\theta }_2}{2}\right)\ } \\ 
cos{(\theta }_1/2){\mathrm{cos} \left(\frac{{\theta }_2}{2}\right)\ } \end{array}
 &  \begin{array}{c}
 \begin{array}{cc}
e^{i{\varphi }_1}\mathrm{sin}\mathrm{}({\theta }_1/2)e^{-i{\varphi }_2}\mathrm{sin}\mathrm{}({\theta }_2/2) & -cos{(\theta }_1/2)\mathrm{cos}\mathrm{}({\theta }_2/2) \end{array}
 \\ 
 \begin{array}{cc}
-e^{i{\varphi }_1}\mathrm{sin}\mathrm{}({\theta }_1/2)\mathrm{cos}\mathrm{}({\theta }_2/2) & -cos{(\theta }_1/2)e^{i{\varphi }_2}\mathrm{sin}\mathrm{}({\theta }_2/2) \end{array}
 \end{array}
 \end{array}
\right) 
\end{equation} 
\end{widetext}

\noindent       However, the eigenfunction ${\psi }_3$ includes the eigenfunction of electrons which belongs to Mo-side with energy $E_1$ , and the eigenfunction of electrons which belongs to W-side with energy $({-E}_2+{\lambda }_2)$. The eigenvalue related to ${\psi }_3$ is ($E_1-(-E_2+{\lambda }_2)$). \\

Up to now, we assumed that the electrons and holes moves freely in the conduction and valence band, respectively. But the electron and hole of exciton are coupled to each other. In the other words, the coupling and the lattice potentials should be added to the total Hamiltonian martix at the begining of the calculation. After adding the all potential terms, the $4\times4$ Hamiltonian matrix of Eq.6 can be divided to four $2\times2$ matrices.  It has been shown that if the non-diagonal elements of the $2\times2$ matrix, which is palced at the down right corner of $4\times4$ Hamiltonian matrix, be equal to zero and the low momentum behavior are only considered the next calculations can be simplified\cite{R30}. Under these assumptions, they have shown that Eq.13 is satisfied. Instead of using the assumptions, we use a mathematical trick i.e., we consider the free electrons and holes and found the eigenvalue of the eigenfunction ${\psi }_3$. Now, if we add the lattice potential $V_I=V_e+V_h$  and Coulomb potential $V_C$ to the eigenvalue of ${\psi }_3$ we will find the energy equation of the exciton as below
\begin{equation} \label{GrindEQ__9_} 
E_{exc}=E^{e,\uparrow }_{Mo}-E^{e,\uparrow }_W+V_I+V_C=E^{e,\uparrow }_{Mo}+E^{h,\downarrow }_W+V_I+V_C 
\end{equation} 
We can fit a parabola to the energy dispersion curve of Mo-side and W-side near $K\ (K^{\prime})$-point and show (Appendix A)  $E^{e,\uparrow }_{Mo}\approx \frac{{\hslash }^2k^2_1}{2m_1}+\frac{{\Delta }_1}{2}$ and $E^{e,\uparrow }_W\approx -\frac{{\hslash }^2k^2_2}{2m_2}-\frac{{\Delta }_2}{2}+{\lambda }_2$. Therefore, the energy dispersion relation of exciton is as below
\begin{equation} \label{GrindEQ__10_} 
E_{exc}=\frac{{\hslash }^2k^2_1}{2m_1}+\frac{{\hslash }^2k^2_2}{2m_2}+\frac{{\Delta }_1+{\Delta }_2}{2}-{\lambda }_2+V_I+V_C 
\end{equation} 
It means that the final result of our mathematical trick is equal to the final result of Li et al.\cite{R30}. In the otherwords, we used a simpler method and showed that the final results are same.

\noindent      Now a question can be asked. How it would be in other kinds of interfaces? The effect of interface structure can be understood by adding a deformation gauge field to the Hamiltonian \cite{R32}. An in- plane gauge field, $\overrightarrow{A}=(A_x\left(x,y\right),A_y\left(x,y\right))$, creates a magnetic filed $\overrightarrow{B}=B_z\hat{z}$ which acts as a pseudo-spin-orbit coupling and splits the CBM and VBM and creates the surface states (Appendix A). For example, if $\overrightarrow{A}=(A_x\left(y\right),0)$ for a zigzag interface, then $B_z\neq 0$ and the energy of the surface states located in the vicinity of the interface reads \cite{R32,R34}:
\begin{equation} \label{GrindEQ__11_} 
E=\frac{v_Fp_x}{\mathrm{cosh}\mathrm{}[\frac{1}{\hslash }\int{dy\ A_x(y)]}} 
\end{equation} 
    Therefore, The eigenvalues of zigzag interface differ from armcair interface due to the existance of the surface states. It means that, as it is shown in Appendix A, by changing the geometrical structure of the interface, the eigenvalues of the Hamiltonian, $H$, changes, and it is expected that the exciton binding energy changes, too (Appendix A). It should be noted that the above results can be used for studying the excitons in van der Waals heterostructures of TMDs because the electrons are localized on the top (bottom) layer while the holes are localized on the bottom (top) layer. Ofcourse, the suitable $V_I$ and $V_C$ should be used \cite{R31}. Also, for studying the bright exciton, one should only use the Hamiltonian $E^{h,\uparrow }_W=-E^{e,\downarrow }_W$ instead of $E^{h,\downarrow }_W=-E^{e,\uparrow }_W$.

\noindent 

\section{Numericall calculations}

\noindent

\noindent         By using the Fourier transformation, one can find the real-space version of Eq.10. Since, the center-of-mass and relative space coordinates are 
\begin{equation} \label{GrindEQ__12_} 
\overrightarrow{R}(X,Y)=\frac{1}{M}\left(m_e{\overrightarrow{r}}_e+m_h{\overrightarrow{r}}_h\right) 
\end{equation} 
\begin{equation} \label{GrindEQ__13_} 
\overrightarrow{r}(x,y)={\overrightarrow{r}}_e-{\overrightarrow{r}}_h 
\end{equation} 
the Hamiltonian of exciton in real space will be equal to
\begin{equation} \label{GrindEQ__14_} 
H=-\frac{{\hslash }^2}{2M}{\mathrm{\nabla }}^2_R-\frac{{\hslash }^2}{2\mu }{\mathrm{\nabla }}^2_r+V_C\left(r\right)+V_I(\overrightarrow{R},\overrightarrow{r}) 
\end{equation} 
where, $M=m_e+m_h$ and $\mu =\frac{m_em_h}{m_e+m_h}$ . By considering the symmetry of $V_I$ and by using the Born-Oppenheimer approximation (BOA) it can be shown that the corresponding Schrodinger's equations for the relative motion and center-of-mass motion read \cite{R33}
\begin{equation} \label{GrindEQ__15_} 
\left[-\frac{{\hslash }^2}{2\mu }{\mathrm{\nabla }}^2_r+V_C\left(r\right)+V_I\left(X,\overrightarrow{r}\right)\right]\mathrm{\Theta }\left(X,r\right)=E(X)\mathrm{\Theta }\left(X,r\right) 
\end{equation} 
\begin{equation} \label{GrindEQ__16_} 
\left[-\frac{{\hslash }^2}{2M}{\mathrm{\nabla }}^2_X+E\left(X\right)\right]\mathrm{\Psi }\left(X\right)=E_{gr}\mathrm{\Psi }\left(X\right) 
\end{equation} 
where, $E_{gr}$ is the ground state energy. By discretizing Eq.18, one can find the minimum eigenvalue $E\left(X\right)$ under open boundary conditions for both directions. By using the minimum value, the exciton energy, $E_{gr}$ , can be found by solving the Eq.19. The binding energy will be found by using the relation ${E_{bi}=E}_f-E_{gr}$ where $E_f$ is the energy of a noninteracting electron-hole pair at the interface.

\noindent    How can one discretize Eq.18?  Liu et al., introduced a three-band tight-binding model for describing the low-energy physics in a monolayer of TMDs \cite{R35}. They showned that the conduction and valence bands are accurately described by $d$ orbitals of metal atoms. Their model involving up to the third-nearest-neighbor hoppings can well reproduce the energy band in the entire Brillouin zone \cite{R35}. Therefore, we can assume that the electrons and holes hop between metal atoms that construct a triangular lattice structure. It means that the relative coordinate, $\overrightarrow{r}$, moves on a triangular lattice structure when the electron (hole) the hop and hole (electron) is fixed. Therefore, we consider the triangular lattice of metal atoms and discretize the Eq.18.  It can be shown that the hopping integral on triangular lattice is equal to $\frac{{\hslash }^2}{3\mu a^2}$, where $a$ is the lattice constant (Appendix B). It shoul be noted that a nanoribbon can be constructed by repeating a supercell. In electron-side and hole-side of the nanoribbon, we consider a supercell. The geometry and the number of atoms in each supercell change by changing the geometry of the interface. In consequence, in tight-binding method, the elements of the Hamiltonian matrix changes by changing the geometry of the interface and the boundary conditions are implimented.

\noindent     It is expected that the excitons are created in the vicinity of the interface line by competition between Coulomb and band offset potentials \cite{R32, R33}. Therefore, we can consider  $\mathrm{\Psi }\left(X\right)\propto e^{-\alpha \left|X\right|}$, where $\alpha =\frac{\sqrt{2M(E\left(X\right)-E_{gr})}}{\hslash }>0$ and$\ X$ is the distance from the interface line in $\hat{x}$-direction. It means that $E_{gr}<$ $E\left(X\right)$ because $E\left(X\right)<0$. But, $E\left(X\right)$ depends on the competition between  $V_I$ and $V_C$, and in consequence, the broadening of $\mathrm{\Psi }\left(X\right)$ and its value depend on the competition. Hence, the values $V_I$ and $V_C$ are very important for calculating $E\left(X\right)$ and $\mathrm{\Psi }\left(X\right)$. 

\noindent However, what are the suitable formulas of  $V_I$ and $V_C$ for the numerical calculations? The Coulomb potential and its usage for studying the Hydrogen-like atoms and the dielectric properties of two-dimensional materials have been widely studied \cite{R31,R36,R37,R38,R39,R40,R41,R42,R43,R44}. Felbacq et al., have used the below formula for studying the dielectric properties of two-dimensional materials \cite{R42}(Appendix C):
\begin{equation} \label{GrindEQ__17_} 
V_C=-(\frac{1}{\sqrt{r^2+\sqrt{\tau }}}+\frac{\tau }{r}) 
\end{equation} 
where, $\tau =\left(\frac{8}{3}\right){\times 10}^{-4}$. Also, they showed that their results are in a good agreement with the results of others \cite{R3}. Therefore, we use the above equation in the following numerical calculations and, for $r=0$ we set  $V_C=U_0=cte$. The assupmtion $V_C(r=0)=U_0=cte$ have been used by Ref.32 and Ref.42 in order to make the calculation convergent. \\

The lattice potential $V_I$ possesses the translational symmetry along the width of the nanoribbon and changes along its length in type II lateral heterostructure of TMDs. Also, the conduction and valence band edges as functions of position are regarded as the step functions \cite{R33}. Lau et al., has numerically modeled the interface potential as $V_e=\frac{V_0+\delta}{2}$${(1-tanh(\frac{x_e}{w}))}$ and 
$V_h=-\frac{V_0-\delta}{2}$${(1-tanh(\frac{x_h}{w}))}$\cite{R33}. Here, $w$ is the width of the interface which is very small in sharp interface and $\delta$ characterizes the difference of the band offsets for electron and hole. they have considered the symmetric heterostructures with $\delta=0$ and $w=0.1a$ where $a$ is the lattice constant\cite{R33}. We use a model such that it covers all the length of the nanoribbon. Therefore by sssuming a symmetric heterostructure with type- II interface and finite width, we will use the below formula for lattice potential in the next calculations.
\begin{equation} \label{GrindEQ__18_} 
V_I=\frac{V_0}{2Max(V_I)}(1-{\mathrm{tanh} \left(\frac{x}{w}\right)\ }) 
\end{equation} 
where, $w$ is the width of the interface,  $x$ is x-coordinate on triangular lattice, $V_0$  is the band offset voltage, 

\noindent and $Max(V_I)$ is the maximum value of $V_I$. It should be noted that for sharp interfaces, $w$ is very small.

\noindent

\section{Results and discussion}

\noindent

\noindent         Theoretically, in Eq.10 for $k_1=k_2=0$, if  $E_g=\frac{{\Delta }_1+{\Delta }_2}{2}-{\lambda }_2\ $, we can write
\begin{equation} \label{GrindEQ__19_} 
E_g-E_{exc}={-(V}_I+V_C) 
\end{equation} 
     where, ${(E}_g-E_{exc})\ $is the binding energy of exciton. The Coulomb potential attempts to bind the electron and hole while the lattice potential at interface attempts to separate them and prefers to place them on the complementary sides of the interface. Therefore, the properties of the exciton depend on the competition between these potentials, especially at the interface. But, the plane integrated modular square wave function of VBM and CBM for different LHSs are localized on W-side and Mo-side, respectively \cite{R32}.  Hence, as the Coulomb potential decays rapidly from the interface line, it is expected that the excitons are created in the vicinity of the interface. Ofcourse, Kang et al., investigated the band offsets and heterostructures of monolayer and few-layer transition-metal dichalcogenides MX${}_{2}$ (M=Mo, W; X=S, Se, Te) and showed that $V_0\approx 0.26\ $eV\cite{R18}. A tight-binding model has been introduced for studying the properties of the lateral heterostructures of two-dimensional materials \cite{R46}. They demonstrtated that the onsite energy ${\Delta }_2\to {\Delta }_2+0.26$ in the presence of ${\Delta }_1$. So, $E_g=1.77\ $($1.57$) eV for MoS${}_{2}$-WS${}_{2}$ (MoSe${}_{2}$-WSe${}_{2}$) lateral heterostructure because,  ${\Delta }_{Mos_2}=1.66$ eV, ${\Delta }_{Ws_2}=1.79$ eV, ${\Delta }_{M{Se}_2}=1.47$ eV, ${\Delta }_{W{Se}_2}=1.60$ eV,  ${\lambda }_{WS_2}=0.215$ eV, and ${\lambda }_{W{Se}_2}=0.23$ eV \cite{R46}.  

\noindent     Now let us find the binding energy of exciton numerically.  Fig.1 shows the triangular grid of a lateral heterostructure of 2D-TMDs with an armchair interface. If the scale of $x(y)$-axis is multiplied by $\sqrt{3}a/2$ ($a/2$), the structure of the zigzag interface will be found.

\begin{figure}[]
\includegraphics[width=\columnwidth]{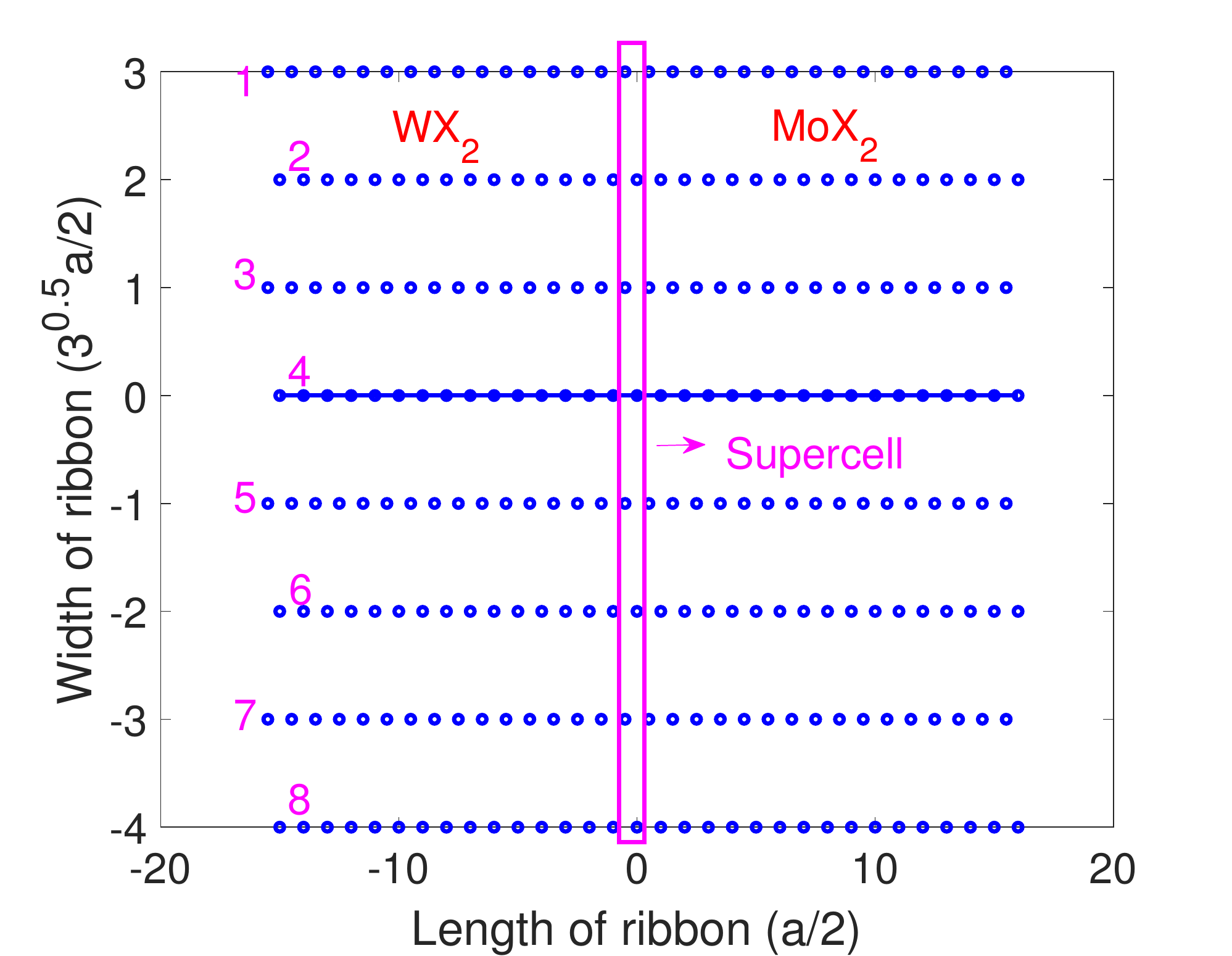}
\caption{\label{fig:epsart} (Color online) Triangular lattice structure of a lateral heterostructure with an armchair interface.}
\end{figure}

\noindent     First, we consider the armchair interface. Here, the hopping integral $\frac{{\hslash }^2}{3\mu a^2}$ is equal to 0.0218 eV because $m_{e(h)}=0.32\ m_0$ and $a=3.325$ Angstrom ($A^0$) \cite{R33}. In the type-II heterostructures, the interface is atomically sharp, and in consequence, we can set $w=0.003$ $A^0$ in Eq.21. Because, for $r=0$ the value of $V_C$ from Eq.20 will be at the order of ${10}^4$, we set $V_C(r=0)=-1.5\ $eV and will show how its effects on the final results can be compensated by a suitable choice of band offset energy, $V_0$ . Fig.2 shows a typical comparison between  $V_I$  and $V_C$. It should be noted that in each supercell, there are eight atoms, and four of them have the same x-coordinate, and consequently, the same lattice potential $V_I$.

\begin{figure}[t!]
\includegraphics[width=\columnwidth]{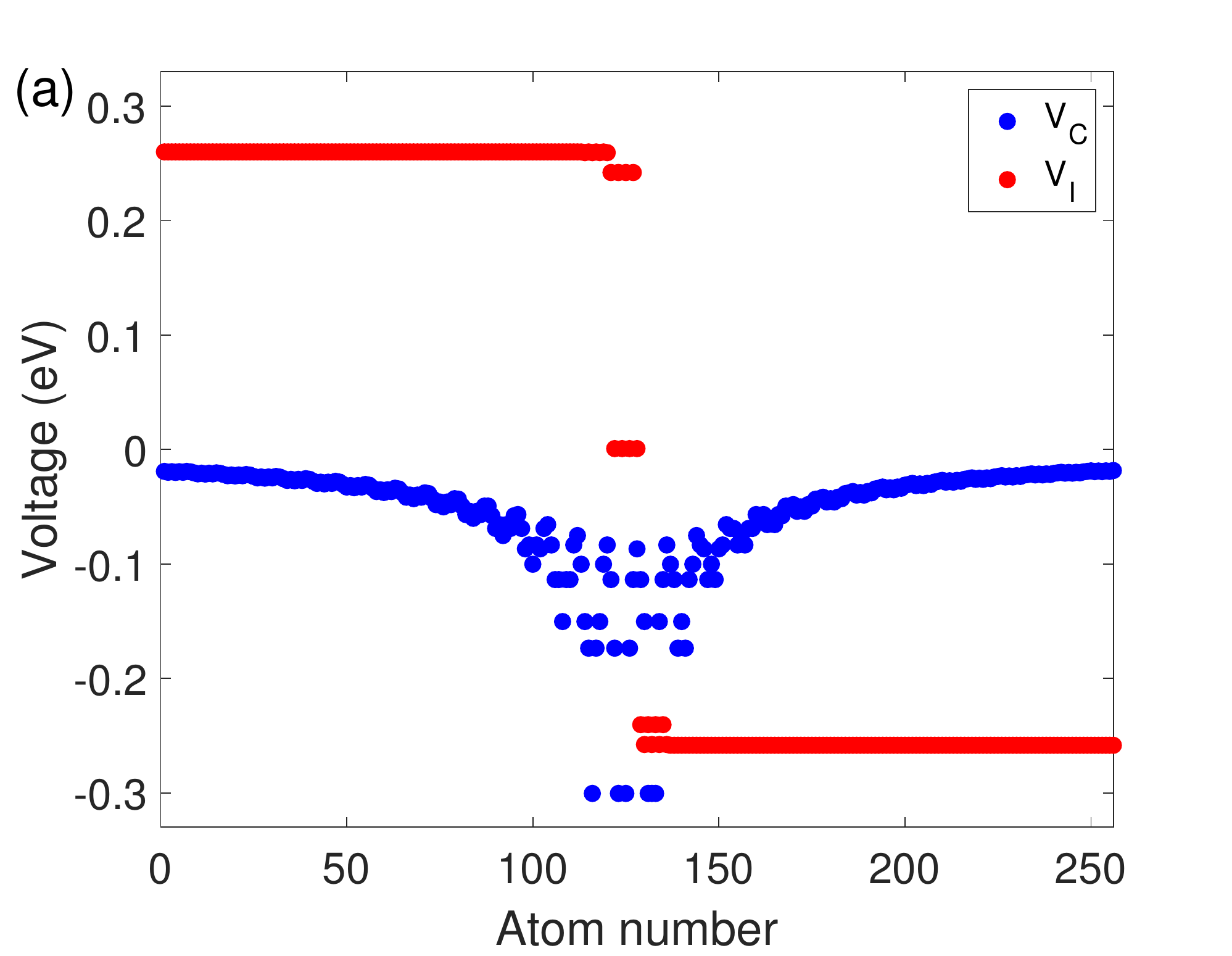}
\includegraphics[width=\columnwidth]{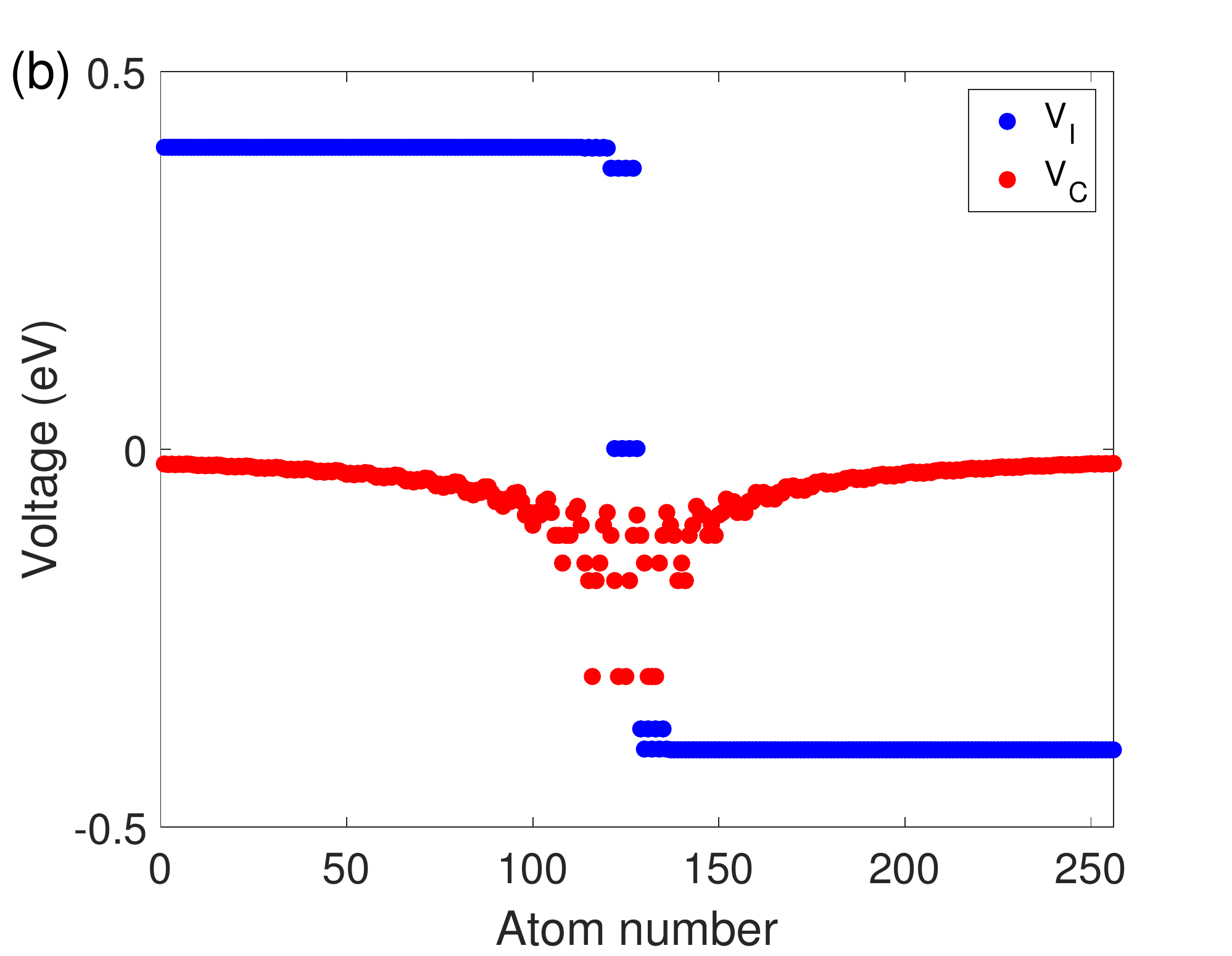}
\caption{\label{fig:epsart}(Color online) Lattice and Coulomb potentials.}${{\mathrm{V}}}_{{\mathrm{C}}}\left({\mathrm{r}}{\mathrm{=}}{\mathrm{0}}\right){\mathrm{=-}}{\mathrm{1}}.{\mathrm{5}}{\mathrm{\ }}{\mathrm{eV}}$\text{ (not shown).}${(a){V}}_{0}=0.26$\text{ eV.} ${and (b){V}}_{0}=0.4$\text{ eV.}
\end{figure}

\noindent      As Fig.2 shows the Coulomb potential has a significant value only at the interface and couples the electrons and holes at the region. For studying the effect of  ${\mathrm{V}}_{\mathrm{C}}$, we should obtain $E(X)$ for different values of  ${\mathrm{V}}_{\mathrm{C}}$ by attention to the value of the band offset voltage.  It has been shown that there are two characteristic behaviors, the regime of small band offset ($V_0<0.1$ eV) and large band offset ($V_0>0.4$ eV) \cite{R33}. For sufficiently large $V_0$, the electron and hole are well separated into opposite regions, while for small and intermediate $V_0$, the separation is weak, and in consequence, on-site Coulomb interaction plays the main role. For example, it has been shown that for $V_0=0.2$ eV, the binding energy of the interface exciton is about 0.22 eV which is about 0.1 eV smaller than that of the 2D exciton \cite{R33}. The effect of Coulomb potential on the energy of excitons ($E(X)$) is shown in Fig.3, for $V_0=0.26$ eV. It can be seen that the exciton energy depends on ${\mathrm{V}}_{\mathrm{C}}\left(\mathrm{r=0}\right)$ because $\left|V_C\right|>\left|V_I\right|$ at some atomic sites near the interface (Fig.2(a)). However, for $V_0=0.4$ eV, as Fig.4 shows, the second minimum eigenvalue does not change when the value of ${\mathrm{V}}_{\mathrm{C}}\left(\mathrm{r=0}\right)$ changes. Under this condition, $\left|V_C\right|<\left|V_I\right|$ and lattice potential dominates (Fig.2(b)), and therefore, the effect of ${\mathrm{V}}_{\mathrm{C}}\left(\mathrm{r=0}\right)$ on the second minimum of eigenvalue is negligible. We consider $V_0=0.4$ eV and the second minimum eigenvalue as  $E(X)$ in the next calculations and will show that under these conditions the correct binding energy can be calculated for MoX${}_{2}$-WX${}_{2}$ LHSs. Also, as Fig.4 shows, the Coulomb potential decreases the second minimum of $E(X)$  energy by 0.26 eV. It means that the system is more stable now.  As the first approximation, we can consider the difference as the binding energy of exciton which is of the same order as a two-dimensional exciton in TMDs. In what follows, we will show that under which conditions the approximation is satisfied.

\begin{figure}[h]
\includegraphics[width=\columnwidth]{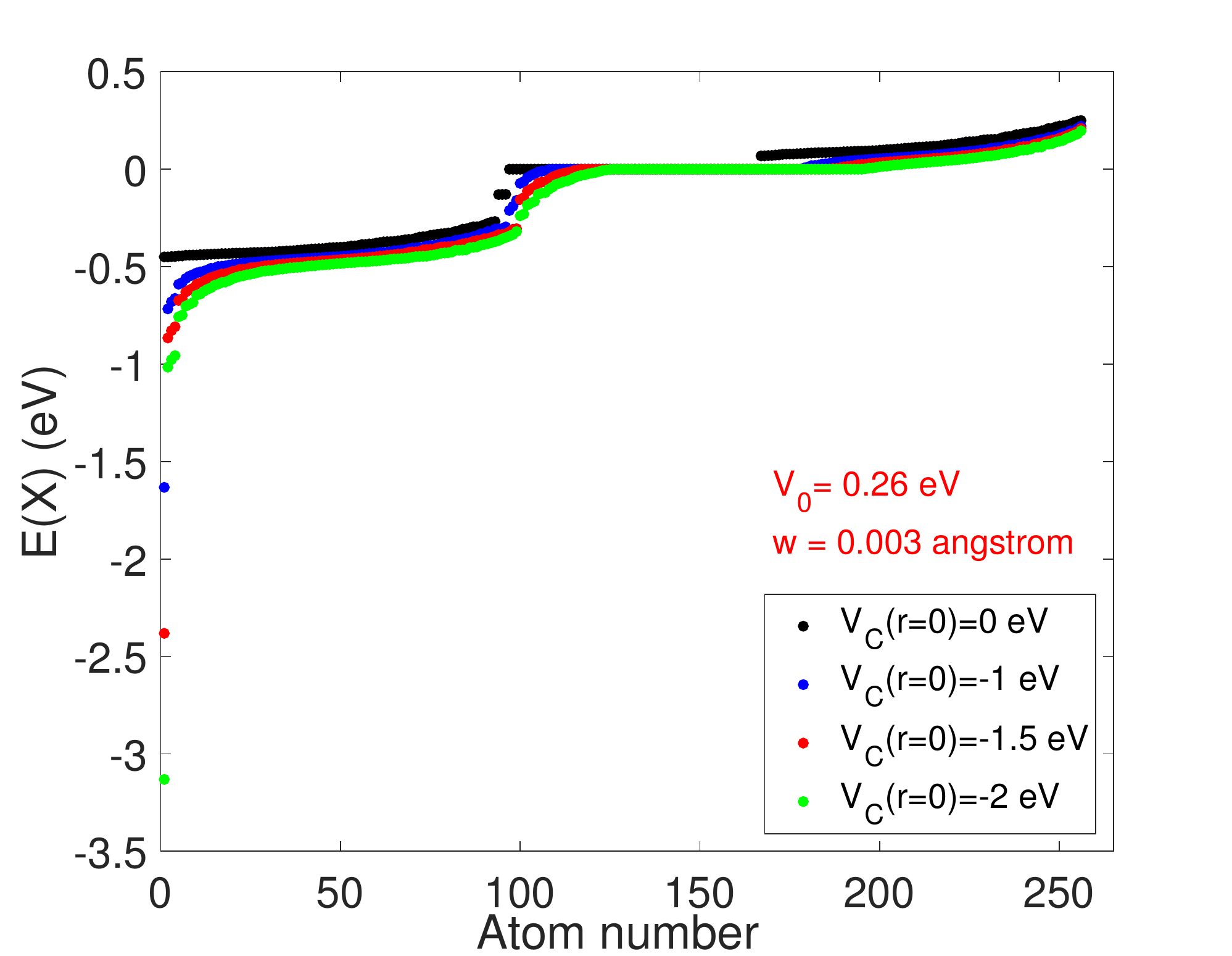}
\caption{\label{fig:epsart}(Color online) The effect of Coulomb potential on the energy of exciton when}${{V}}_{0}=0.26$\text{ eV.}
\end{figure}

\begin{figure}[h]
\includegraphics[width=\columnwidth]{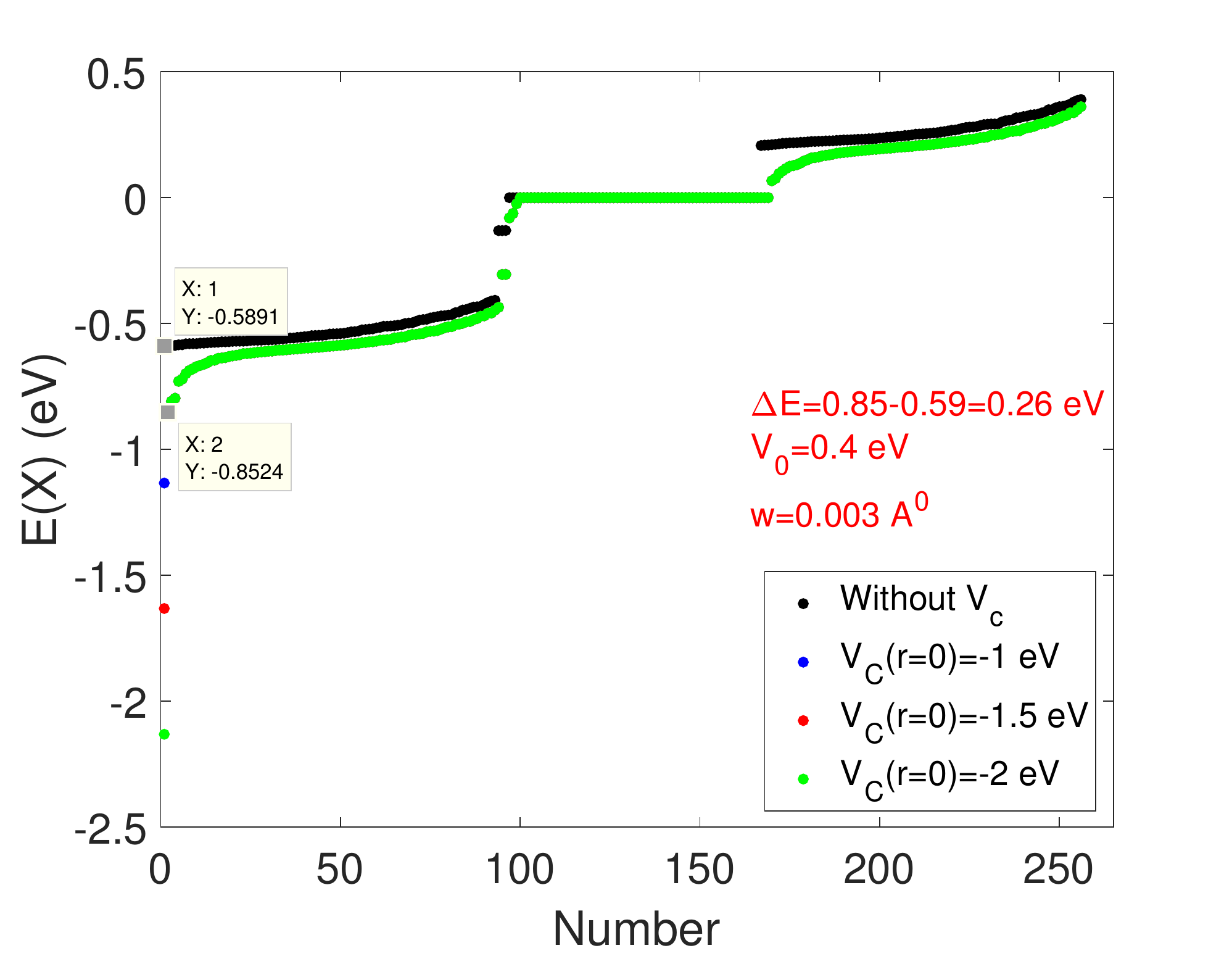}
\caption{\label{fig:epsart}(Color online) The effect of Coulomb potential on the energy of exciton when}${{V}}_{0}=0.4$\text{ eV.}
\end{figure}

\noindent 

\noindent  Our guess for the wave function of center-of-mass was $\mathrm{\Psi }\left(X\right)\propto e^{-\alpha \left|X\right|}$ where $\alpha =\frac{\sqrt{2M(E\left(X\right)-E_{gr})}}{\hslash }>0$. As electron and hole are well separated into opposite regions for $V_0=0.4$ eV even for very long nanoribbon \cite{R1,R5} and considering the behavior and value of the Coulomb potential compared to the lattice potential (Fig.2(b)), it is expected that $\mathrm{\Psi }\left(X\right)$ decays to $1/e$ of its maximum value for a specific value of $X$ (called $X_b)$. For $X=X_b$ we have  $\frac{\sqrt{2M(E\left(X\right)-E_{gr})}}{\hslash }=\frac{1}{X_b}$ and in consequence:
\begin{equation} \label{GrindEQ__20_} 
E_{gr}=E\left(X\right)-\frac{{\hslash }^2}{2MX^2_b} 
\end{equation} 
Therefore, by increasing $X_b$ (well separation of electron and hole) the term $\frac{{\hslash }^2}{2MX^2_b}$ decreases rapidly and $E_{gr}\to E\left(X\right)$. But, the minimum value of  $X^{min}_b\ $is equal to $\frac{m_ha}{2M}=\frac{a}{4}=0.83$ Angstrom when one electron is at interface ($x=0$) and one hole is at ($x=a/2$). So, the maximum value of $\frac{{\hslash }^2}{2MX^2_b}=0.08$ eV and $E_{gr}=E\left(X\right)-0.08$ . Under this condition, $E_{bi}=0.34$ eV which is equal to the binding energy of 2D exciton, approximately \cite{R33}. As a result, it can be concluded that near the interface $E_{bi}=0.34$ eV and far from it $E_{bi}=0.26$ eV. It means that the binding energy of exciton depends on its distance from the interface.

\noindent     Now let us, study the effect of the interface structure on binding energy. Fig. 5 shows the comparison between the binding energy of zigzag and armchair interfaces. As it shows, the difference is $\Delta E_{min}\left(X\right)=0.34$ meV. $\Delta E_{min}\left(X\right)$ is created by the deformation gauge field which is a pseudo spin-orbit coupling. Its value is small due to its nature.

\begin{figure}[bt]
\includegraphics[width=\columnwidth]{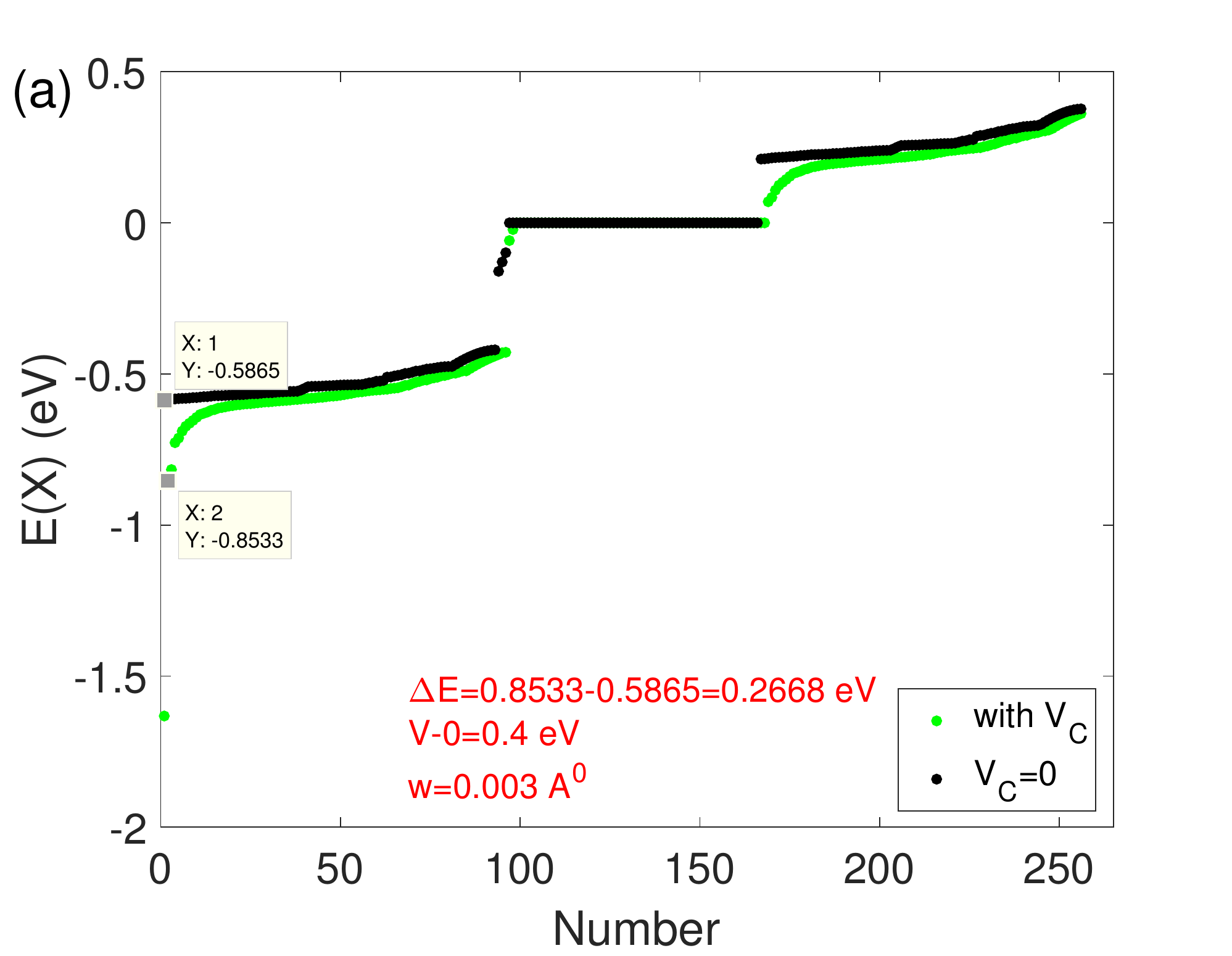}
\includegraphics[width=\columnwidth]{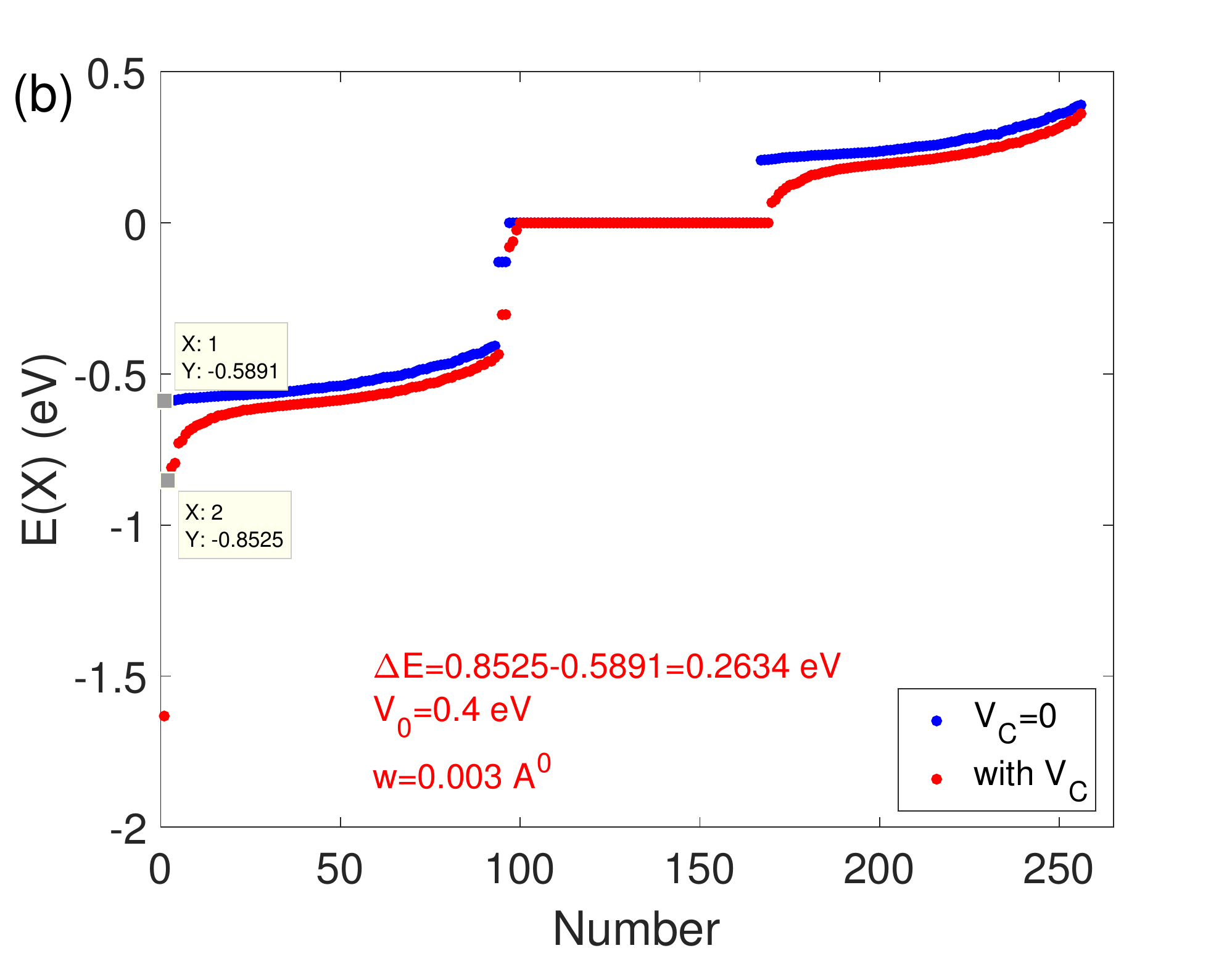}
\caption{\label{fig:epsart}Fig.5 (Color online) Comparison between binding energy of (a) zigzag and (b) armchair interfaces.}${{V}}_{0}=0.4$\text{ eV.}
\end{figure}

\section{Summary}

\noindent

\noindent     We studied the dark exciton in two --dimensional dichalcogenide LHSs with a sharp interface. We introduced a low-energy effective Hamiltonian model and found the energy dispersion relation of exciton and showed how it depends on the onsite energy of composed materials and their spin-orbit coupling strengths. It was shown that the balance between the Coulomb and offset potential implies the behavior of exciton, especially at the interface. Also, by assigning a deformation gauge field to the geometrical structure of the interface, we  could find the effect of the geometry on the binding energy of exciton. By discretization of the real-space version of dispersion relation on a triangular lattice, the exciton binding energy calculated as 0.34 eV near (far from) the interface i.e., the binding energy of exciton depends on its distance from the interface line. Finally, we could show that the binding energy of a zigzag interface increases by 0.34 meV in comparison with an armchair interface due to the pseudo-spin-orbit coupling term (deformation gauge field). The results can be useful in the design of new optoelectronic devices with improved performance and characteristics.  

\section{Data availablity}

\noindent

\noindent     The data that supports the findings of this study are available within the article and its Appendices.

\appendix

\section{}


\noindent      The low-energy two-band effective Hamiltonian of spin-up electrons near the K-point is given by \cite{R46}
\begin{equation}
H=\left( \begin{array}{cc}
\frac{\Delta }{2} & atk_- \\ 
atk_+ & -\frac{\Delta }{2}+\lambda  \end{array}
\right)                                                                                                                                                                                                                   
\end{equation}

\noindent where, $k_{\pm }=k_x\pm k_y$, and $\Delta $ , $a$, and $t$ are the energy band gap, lattice constant, and hopping integral, respectively. Here, $2\lambda $ is the spin-orbit coupling (SOC) strength. It can be shown that the eigenvalues are given by
\begin{equation}
E=\frac{\lambda }{2}\pm \sqrt{{\left(\frac{\Delta }{2}-\frac{\lambda }{2}\right)}^2+a^2t^2k^2}                                                                                                                
\end{equation}

\noindent By defining $\mathrm{cos}\mathrm{}(\frac{\theta }{2})$=$atk/\sqrt{(E-\frac{\mathrm{\Delta }}{2})(2E-\lambda )}$ and $={tan}^{-1}(\frac{k_y}{k_x})$ , one can easily diagonalize the Hamiltonian, $H$,  by  using the matrix $P=\left( \begin{array}{cc}
\mathrm{cos}\mathrm{}(\frac{\theta }{2}) & e^{-i\varphi }\mathrm{sin}\mathrm{}(\frac{\theta }{2}) \\ 
e^{i\varphi }\mathrm{sin}\mathrm{}(\frac{\theta }{2}) & \mathrm{-}\mathrm{cos}\mathrm{}(\frac{\theta }{2}) \end{array}
\right)$ and shows
\begin{equation}
P^{-1}H P=\left( \begin{array}{cc}
E & 0 \\ 
0 & -E+\lambda  \end{array}
\right)                                                                                                                                                                                                                    
\end{equation}

\noindent For $k=0$ , the energy eigenvalues are $E=\frac{\lambda }{2}\pm \left(\frac{\Delta }{2}-\frac{\lambda }{2}\right)$ and in consequence the conduction band minimum  (CBM) and valence band maximum (VBM) are equal to $\frac{\Delta }{2}$ and $-\frac{\Delta }{2}+\lambda $, respectively. For spin-down electrons, CBM and VBM are $\frac{\Delta }{2}$ and $-\frac{\Delta }{2}-\lambda $, respectively. So, the band splitting happens at VBM by the SOC.

\noindent     If the eigenfunctions ${\phi }_1={\left({\mathrm{cos} \left(\frac{\theta }{2}\right)\ \ }e^{i\varphi }\mathrm{sin}\mathrm{}(\frac{\theta }{2})\right)}^T$ and ${\phi }_2={\left(e^{-i\varphi }{\mathrm{sin} \left(\frac{\theta }{2}\right)\ {\mathrm{-}\mathrm{cos} \left(\frac{\theta }{2}\right)\ \ }\ }\right)}^T$are used it can be shown that

\begin{equation}
 H{\phi }_1=E{\phi }_1 
\end{equation}
\begin{equation}
H{\phi }_2=(-E+\lambda ){\phi }_2. 
\end{equation}

\noindent     For adding a gauge field, $\overrightarrow{A}$, to the Hamiltonian, we should use the covariant derivatives. It means that we should replace $\overrightarrow{p}$ by $\overrightarrow{p}-\overrightarrow{A}$ in the Hamiltonian matrix. If we consider the equation $H^2\psi =E^2\psi $, it can be shown that
\begin{equation}
E=\frac{\lambda }{2}\pm \sqrt{{\left(\frac{\Delta }{2}-\frac{\lambda }{2}\right)}^2+v^2_F{\left(\overrightarrow{p}-\overrightarrow{A}\right)}^2\pm \hslash v^2_FB_z}                                                                                                                                                                                                  
\end{equation}

\noindent where, $v_F$ is Fermi velocity, $\overrightarrow{p}=\hslash \overrightarrow{k}$ , $\overrightarrow{A}=(A_x\left(x,y\right),A_y\left(x,y\right))$, and $\overrightarrow{B}=\mathrm{\nabla }\times \overrightarrow{A}$. If $k=0$, and by neglecting the term $A^2$, it can be shown that
\begin{equation}
E=\frac{\lambda }{2}\pm \sqrt{{\left(\frac{\Delta }{2}-\frac{\lambda }{2}\right)}^2\pm \hslash v^2_FB_z}                                                                                                                 
\end{equation}

\noindent If $\frac{\Delta -\lambda }{2}\gg \ \hslash v^2_FB_z$ then
\begin{equation}
 E_{1\eqref{GrindEQ__2_}}=\frac{\Delta }{2}+(-)\frac{\hslash v^2_FB_z}{2}
\end{equation}
 and 
\begin{equation}
E_{3\eqref{GrindEQ__4_}}=-\frac{\Delta }{2}+\lambda +(-)\frac{\hslash v^2_FB_z}{2}.
\end{equation}
 Therefore, the term $\hslash v^2_FB_z$ is pseudo-spin orbit coupling and splits not only VBM but also CBM. Because for $E=0$ the Eq.(A-6) has a non-trivial solution, the surface states exist.


\section{}

\noindent      In $xy$-plane, one can define three vectors $\overrightarrow{u}=(u_x,0)$, $\overrightarrow{v}=(v\ cos\alpha ,\ v\ sin\alpha )$, and $\overrightarrow{w}=\left(w\ cos\beta ,\ w\ sin\beta \right)$ and shows that 
\begin{equation}
 {\partial }^2_u={\partial }^2_x
\end{equation}
\begin{equation}
 {\partial }^2_v={(cos}^2\alpha ){\partial }^2_x+\left({sin}^2\alpha \right){\partial }^2_y+2(sin\alpha )(cos\alpha ){\partial }^2_{xy}
\end{equation}
\begin{equation}
{\partial }^2_w={(cos}^2\beta ){\partial }^2_x+\left({sin}^2\beta \right){\partial }^2_y+2(sin\beta )(cos\beta ){\partial }^2_{xy}.
\end{equation}
 In a triangular lattice, $\alpha =60$ and $\beta =120$ and in consequence \cite{R47}
\begin{equation}
{\partial }^2_x+{\partial }^2_y=\frac{2}{3}({\partial }^2_u+{\partial }^2_v+{\partial }^2_w)                                                                                                  
                                                                                                                                                                                                                                \end{equation}

\noindent Now by discretizing the right-hand side of the Eq.(B-4), the below equation can be derived\cite{R47}
\begin{widetext}
\begin{equation}
-\frac{{\hslash }^2}{2m}\left(\frac{{\partial }^2\psi }{\partial x^2}+\frac{{\partial }^2\psi }{\partial y^2}\right)=-\frac{{\hslash }^2}{3ma^2}({-6\psi }_{i,j,k}+{\psi }_{i+1,j,k}{+\psi }_{i-1,j,k}{+\psi }_{i,j+1,k}{+\psi }_{i,j-1,k}{+\psi }_{i,j,k+1}{+\psi }_{i,j,k-1}) 
\end{equation}
\end{widetext}
\noindent where, $i$, $j$, and $k$ are in $\overrightarrow{u}$, $\overrightarrow{v}$, and $\overrightarrow{w}$ directions. Therefore, the hopping integral is equal to  $=-\frac{{\hslash }^2}{3ma^2}$ where, $a$ is the lattice constant.


\section{}

It is assumed that two electric charges $e$ and $e^\prime$ are located at positions $(\rho_{0},z_{0} )$ and $(0,z_{0}^\prime)$, respectively in a slab of dielectric constant $\epsilon_{f}$. The slab width is $d$ and is surrounded by a medium of dielectric constant $\epsilon_{1}$ $(z\leqslant-d/2)$  and a medium of dielectric constant $\epsilon_{2}$$(z\geqslant+d/2)$. Felbacq et al., have shown that the electrostatic energy between both charges is given by\cite{R42}:
\begin{equation}
V(\rho,z_{0},z_{0}^\prime )=\frac{{e}}{(2\pi\epsilon_{f} d)}I(r,x,y)
\end{equation}
 where, $\epsilon_{f}$ is the dielectric constant of slab, $r=\frac{{\rho}}{d}$, $x=\frac{{z_{0}}}{d}$, $y=\frac{{z_{0}^\prime}}{d}$ , and $u=\frac{{k}}{d}$. Here, 
\begin{equation}
I(r,x,y)=\int_a^b  f(u,x,y) J_{0}(r,u) du +\frac{{1}}{2\sqrt{r^2+{\mid x-y\mid}^2}}
\end{equation}
where $f(u,x,y)=W(u,x,y) -\frac{{1}}{2}\exp(-u\mid x-y\mid)$ and $J_{0}(r,u)$ is Bessel function. By proofing a theorm, they have shown that the kernel of the first integral tends exponentially fast toward zero and it defines a function that regular near the origin $x=y=0$. It means that the screened electrostatic potential in a dielectric slab can be considered as the usual Coulomb potential plus a correction term\cite{R42}. If the slab width is large compared to the relative distance between the two electric charges the correction term is exponentially small. Also, if the height z approaches zero, the expression does not present any divergence\cite{R42}. The expression is suitable for further numerical calculations aiming to compute the binding energy of excitons in quasi-2D materials\cite{R42}.

\nocite{*}

\bibliography{apssamp}

\providecommand{\noopsort}[1]{}\providecommand{\singleletter}[1]{#1}%
\begin{thebibliography}{46}%
\makeatletter
\providecommand \@ifxundefined [1]{%
 \@ifx{#1\undefined}
}%
\providecommand \@ifnum [1]{%
 \ifnum #1\expandafter \@firstoftwo
 \else \expandafter \@secondoftwo
 \fi
}%
\providecommand \@ifx [1]{%
 \ifx #1\expandafter \@firstoftwo
 \else \expandafter \@secondoftwo
 \fi
}%
\providecommand \natexlab [1]{#1}%
\providecommand \enquote  [1]{``#1''}%
\providecommand \bibnamefont  [1]{#1}%
\providecommand \bibfnamefont [1]{#1}%
\providecommand \citenamefont [1]{#1}%
\providecommand \href@noop [0]{\@secondoftwo}%
\providecommand \href [0]{\begingroup \@sanitize@url \@href}%
\providecommand \@href[1]{\@@startlink{#1}\@@href}%
\providecommand \@@href[1]{\endgroup#1\@@endlink}%
\providecommand \@sanitize@url [0]{\catcode `\\12\catcode `\$12\catcode
  `\&12\catcode `\#12\catcode `\^12\catcode `\_12\catcode `\%12\relax}%
\providecommand \@@startlink[1]{}%
\providecommand \@@endlink[0]{}%
\providecommand \url  [0]{\begingroup\@sanitize@url \@url }%
\providecommand \@url [1]{\endgroup\@href {#1}{\urlprefix }}%
\providecommand \urlprefix  [0]{URL }%
\providecommand \Eprint [0]{\href }%
\providecommand \doibase [0]{http://dx.doi.org/}%
\providecommand \selectlanguage [0]{\@gobble}%
\providecommand \bibinfo  [0]{\@secondoftwo}%
\providecommand \bibfield  [0]{\@secondoftwo}%
\providecommand \translation [1]{[#1]}%
\providecommand \BibitemOpen [0]{}%
\providecommand \bibitemStop [0]{}%
\providecommand \bibitemNoStop [0]{.\EOS\space}%
\providecommand \EOS [0]{\spacefactor3000\relax}%
\providecommand \BibitemShut  [1]{\csname bibitem#1\endcsname}%
\let\auto@bib@innerbib\@empty
\bibitem [{\citenamefont {Lin}, \citenamefont {Williams},\ and\ \citenamefont
  {Connell}(2010)}]{R1}%
  \BibitemOpen
  \bibfield  {author} {\bibinfo {author} {\bibfnamefont {Y.}~\bibnamefont
  {Lin}}, \bibinfo {author} {\bibfnamefont {T.}~\bibnamefont {Williams}}, \
  and\ \bibinfo {author} {\bibfnamefont {J.}~\bibnamefont {Connell}},\
  }\href@noop {} {\bibfield  {journal} {\bibinfo  {journal} {Phys. Chem.
  Lett.}\ }\textbf {\bibinfo {volume} {1}},\ \bibinfo {pages} {277} (\bibinfo
  {year} {2010})}\BibitemShut {NoStop}%
\bibitem [{\citenamefont {Weng}\ \emph {et~al.}(2016)\citenamefont {Weng},
  \citenamefont {Wang}, \citenamefont {Wang}, \citenamefont {Bando},\ and\
  \citenamefont {Golberg}}]{R2}%
  \BibitemOpen
  \bibfield  {author} {\bibinfo {author} {\bibfnamefont {Q.}~\bibnamefont
  {Weng}}, \bibinfo {author} {\bibfnamefont {X.}~\bibnamefont {Wang}}, \bibinfo
  {author} {\bibfnamefont {X.}~\bibnamefont {Wang}}, \bibinfo {author}
  {\bibfnamefont {Y.}~\bibnamefont {Bando}}, \ and\ \bibinfo {author}
  {\bibfnamefont {D.}~\bibnamefont {Golberg}},\ }\href@noop {} {\bibfield
  {journal} {\bibinfo  {journal} {Chem. Soc. Rev.}\ }\textbf {\bibinfo {volume}
  {45}},\ \bibinfo {pages} {3989} (\bibinfo {year} {2016})}\BibitemShut
  {NoStop}%
\bibitem [{\citenamefont {Lv}\ \emph {et~al.}(2015)\citenamefont {Lv},
  \citenamefont {Robinson}, \citenamefont {Schaak}, \citenamefont {Sun},
  \citenamefont {Sun}, \citenamefont {Mallouk},\ and\ \citenamefont
  {Terrones}}]{R3}%
  \BibitemOpen
  \bibfield  {author} {\bibinfo {author} {\bibfnamefont {R.}~\bibnamefont
  {Lv}}, \bibinfo {author} {\bibfnamefont {J.}~\bibnamefont {Robinson}},
  \bibinfo {author} {\bibfnamefont {R.}~\bibnamefont {Schaak}}, \bibinfo
  {author} {\bibfnamefont {D.}~\bibnamefont {Sun}}, \bibinfo {author}
  {\bibfnamefont {Y.}~\bibnamefont {Sun}}, \bibinfo {author} {\bibfnamefont
  {T.}~\bibnamefont {Mallouk}}, \ and\ \bibinfo {author} {\bibfnamefont
  {M.}~\bibnamefont {Terrones}},\ }\href@noop {} {\bibfield  {journal}
  {\bibinfo  {journal} {Acc. Chem. Res.}\ }\textbf {\bibinfo {volume} {48}},\
  \bibinfo {pages} {56} (\bibinfo {year} {2015})}\BibitemShut {NoStop}%
\bibitem [{\citenamefont {Tan}\ and\ \citenamefont {Zhang}(2015)}]{R4}%
  \BibitemOpen
  \bibfield  {author} {\bibinfo {author} {\bibfnamefont {C.}~\bibnamefont
  {Tan}}\ and\ \bibinfo {author} {\bibfnamefont {H.}~\bibnamefont {Zhang}},\
  }\href@noop {} {\bibfield  {journal} {\bibinfo  {journal} {Chem. Soc. Rev.}\
  }\textbf {\bibinfo {volume} {44}},\ \bibinfo {pages} {2713} (\bibinfo {year}
  {2015})}\BibitemShut {NoStop}%
\bibitem [{\citenamefont {Liu}\ \emph {et~al.}(2015)\citenamefont {Liu},
  \citenamefont {Du}, \citenamefont {Deng},\ and\ \citenamefont {Ye}}]{R5}%
  \BibitemOpen
  \bibfield  {author} {\bibinfo {author} {\bibfnamefont {H.}~\bibnamefont
  {Liu}}, \bibinfo {author} {\bibfnamefont {Y.}~\bibnamefont {Du}}, \bibinfo
  {author} {\bibfnamefont {Y.}~\bibnamefont {Deng}}, \ and\ \bibinfo {author}
  {\bibfnamefont {P.}~\bibnamefont {Ye}},\ }\href@noop {} {\bibfield  {journal}
  {\bibinfo  {journal} {Chem. Soc. Rev.}\ }\textbf {\bibinfo {volume} {44}},\
  \bibinfo {pages} {2732} (\bibinfo {year} {2015})}\BibitemShut {NoStop}%
\bibitem [{\citenamefont {Chen}\ \emph {et~al.}(2018)\citenamefont {Chen},
  \citenamefont {Li}, \citenamefont {Chen}, \citenamefont {Ong},\ and\
  \citenamefont {Zhao}}]{R6}%
  \BibitemOpen
  \bibfield  {author} {\bibinfo {author} {\bibfnamefont {P.}~\bibnamefont
  {Chen}}, \bibinfo {author} {\bibfnamefont {N.}~\bibnamefont {Li}}, \bibinfo
  {author} {\bibfnamefont {X.}~\bibnamefont {Chen}}, \bibinfo {author}
  {\bibfnamefont {W.}~\bibnamefont {Ong}}, \ and\ \bibinfo {author}
  {\bibfnamefont {X.}~\bibnamefont {Zhao}},\ }\href@noop {} {\bibfield
  {journal} {\bibinfo  {journal} {2D Mater.}\ }\textbf {\bibinfo {volume}
  {5}},\ \bibinfo {pages} {014002} (\bibinfo {year} {2018})}\BibitemShut
  {NoStop}%
\bibitem [{\citenamefont {Kara}\ \emph {et~al.}(2012)\citenamefont {Kara},
  \citenamefont {Enriquez}, \citenamefont {Seitsonen}, \citenamefont {Voon},
  \citenamefont {Vizzini}, \citenamefont {Aufray},\ and\ \citenamefont
  {Oughaddou}}]{R7}%
  \BibitemOpen
  \bibfield  {author} {\bibinfo {author} {\bibfnamefont {A.}~\bibnamefont
  {Kara}}, \bibinfo {author} {\bibfnamefont {H.}~\bibnamefont {Enriquez}},
  \bibinfo {author} {\bibfnamefont {A.}~\bibnamefont {Seitsonen}}, \bibinfo
  {author} {\bibfnamefont {L.}~\bibnamefont {Voon}}, \bibinfo {author}
  {\bibfnamefont {S.}~\bibnamefont {Vizzini}}, \bibinfo {author} {\bibfnamefont
  {B.}~\bibnamefont {Aufray}}, \ and\ \bibinfo {author} {\bibfnamefont
  {H.}~\bibnamefont {Oughaddou}},\ }\href@noop {} {\bibfield  {journal}
  {\bibinfo  {journal} {Surf. Sci. Rep.}\ }\textbf {\bibinfo {volume} {67}},\
  \bibinfo {pages} {1} (\bibinfo {year} {2012})}\BibitemShut {NoStop}%
\bibitem [{\citenamefont {Duan}\ \emph
  {et~al.}(2014{\natexlab{a}})\citenamefont {Duan}, \citenamefont {Wang},
  \citenamefont {Shaw}, \citenamefont {Cheng}, \citenamefont {Chen},
  \citenamefont {Li}, \citenamefont {Wu}, \citenamefont {Tang}, \citenamefont
  {Zhang}, \citenamefont {Pan}, \citenamefont {Jiang}, \citenamefont {Yu},
  \citenamefont {Huang},\ and\ \citenamefont {Duan}}]{R8}%
  \BibitemOpen
  \bibfield  {author} {\bibinfo {author} {\bibfnamefont {X.}~\bibnamefont
  {Duan}}, \bibinfo {author} {\bibfnamefont {C.}~\bibnamefont {Wang}}, \bibinfo
  {author} {\bibfnamefont {J.~C.}\ \bibnamefont {Shaw}}, \bibinfo {author}
  {\bibfnamefont {R.}~\bibnamefont {Cheng}}, \bibinfo {author} {\bibfnamefont
  {Y.}~\bibnamefont {Chen}}, \bibinfo {author} {\bibfnamefont {H.}~\bibnamefont
  {Li}}, \bibinfo {author} {\bibfnamefont {X.}~\bibnamefont {Wu}}, \bibinfo
  {author} {\bibfnamefont {Y.}~\bibnamefont {Tang}}, \bibinfo {author}
  {\bibfnamefont {Q.}~\bibnamefont {Zhang}}, \bibinfo {author} {\bibfnamefont
  {A.}~\bibnamefont {Pan}}, \bibinfo {author} {\bibfnamefont {J.}~\bibnamefont
  {Jiang}}, \bibinfo {author} {\bibfnamefont {R.}~\bibnamefont {Yu}}, \bibinfo
  {author} {\bibfnamefont {Y.}~\bibnamefont {Huang}}, \ and\ \bibinfo {author}
  {\bibfnamefont {X.}~\bibnamefont {Duan}},\ }\href@noop {} {\bibfield
  {journal} {\bibinfo  {journal} {Nanotechnol.}\ }\textbf {\bibinfo {volume}
  {9}},\ \bibinfo {pages} {1024} (\bibinfo {year}
  {2014}{\natexlab{a}})}\BibitemShut {NoStop}%
\bibitem [{\citenamefont {Heo}\ \emph {et~al.}(2015)\citenamefont {Heo},
  \citenamefont {Sung}, \citenamefont {Jin}, \citenamefont {Ahn}, \citenamefont
  {Kim}, \citenamefont {Lee}, \citenamefont {Cha}, \citenamefont {Choi},\ and\
  \citenamefont {Jo}}]{R9}%
  \BibitemOpen
  \bibfield  {author} {\bibinfo {author} {\bibfnamefont {H.}~\bibnamefont
  {Heo}}, \bibinfo {author} {\bibfnamefont {J.~H.}\ \bibnamefont {Sung}},
  \bibinfo {author} {\bibfnamefont {G.}~\bibnamefont {Jin}}, \bibinfo {author}
  {\bibfnamefont {J.~H.}\ \bibnamefont {Ahn}}, \bibinfo {author} {\bibfnamefont
  {K.}~\bibnamefont {Kim}}, \bibinfo {author} {\bibfnamefont {M.~J.}\
  \bibnamefont {Lee}}, \bibinfo {author} {\bibfnamefont {S.}~\bibnamefont
  {Cha}}, \bibinfo {author} {\bibfnamefont {H.}~\bibnamefont {Choi}}, \ and\
  \bibinfo {author} {\bibfnamefont {M.~H.}\ \bibnamefont {Jo}},\ }\href@noop {}
  {\bibfield  {journal} {\bibinfo  {journal} {Adv. Mater.}\ }\textbf {\bibinfo
  {volume} {27}},\ \bibinfo {pages} {3803} (\bibinfo {year}
  {2015})}\BibitemShut {NoStop}%
\bibitem [{\citenamefont {Gong}\ \emph {et~al.}(2015)\citenamefont {Gong},
  \citenamefont {Lei}, \citenamefont {Ye}, \citenamefont {Li}, \citenamefont
  {He}, \citenamefont {Keyshar}, \citenamefont {Zhang}, \citenamefont {Wang},
  \citenamefont {Lou}, \citenamefont {Liu}, \citenamefont {Vajtai},
  \citenamefont {Zhou},\ and\ \citenamefont {Ajayan}}]{Rten}%
  \BibitemOpen
  \bibfield  {author} {\bibinfo {author} {\bibfnamefont {Y.}~\bibnamefont
  {Gong}}, \bibinfo {author} {\bibfnamefont {S.}~\bibnamefont {Lei}}, \bibinfo
  {author} {\bibfnamefont {G.}~\bibnamefont {Ye}}, \bibinfo {author}
  {\bibfnamefont {B.}~\bibnamefont {Li}}, \bibinfo {author} {\bibfnamefont
  {Y.}~\bibnamefont {He}}, \bibinfo {author} {\bibfnamefont {K.}~\bibnamefont
  {Keyshar}}, \bibinfo {author} {\bibfnamefont {X.}~\bibnamefont {Zhang}},
  \bibinfo {author} {\bibfnamefont {Q.}~\bibnamefont {Wang}}, \bibinfo {author}
  {\bibfnamefont {J.}~\bibnamefont {Lou}}, \bibinfo {author} {\bibfnamefont
  {Z.}~\bibnamefont {Liu}}, \bibinfo {author} {\bibfnamefont {R.}~\bibnamefont
  {Vajtai}}, \bibinfo {author} {\bibfnamefont {W.}~\bibnamefont {Zhou}}, \ and\
  \bibinfo {author} {\bibfnamefont {P.~M.}\ \bibnamefont {Ajayan}},\
  }\href@noop {} {\bibfield  {journal} {\bibinfo  {journal} {Nano Lett.}\
  }\textbf {\bibinfo {volume} {15}},\ \bibinfo {pages} {6135} (\bibinfo {year}
  {2015})}\BibitemShut {NoStop}%
\bibitem [{\citenamefont {Zhang}\ \emph {et~al.}(2017)\citenamefont {Zhang},
  \citenamefont {Chen}, \citenamefont {Duan}, \citenamefont {Zang},
  \citenamefont {Luo},\ and\ \citenamefont {Duan}}]{R11}%
  \BibitemOpen
  \bibfield  {author} {\bibinfo {author} {\bibfnamefont {Z.}~\bibnamefont
  {Zhang}}, \bibinfo {author} {\bibfnamefont {P.}~\bibnamefont {Chen}},
  \bibinfo {author} {\bibfnamefont {X.}~\bibnamefont {Duan}}, \bibinfo {author}
  {\bibfnamefont {K.}~\bibnamefont {Zang}}, \bibinfo {author} {\bibfnamefont
  {J.}~\bibnamefont {Luo}}, \ and\ \bibinfo {author} {\bibfnamefont
  {X.}~\bibnamefont {Duan}},\ }\href@noop {} {\bibfield  {journal} {\bibinfo
  {journal} {Science}\ }\textbf {\bibinfo {volume} {357}},\ \bibinfo {pages}
  {788} (\bibinfo {year} {2017})}\BibitemShut {NoStop}%
\bibitem [{\citenamefont {Huang}\ \emph {et~al.}(2014)\citenamefont {Huang},
  \citenamefont {Wu}, \citenamefont {Sanchez}, \citenamefont {Peters},
  \citenamefont {Beanland}, \citenamefont {Ross}, \citenamefont {Rivera},
  \citenamefont {Yao}, \citenamefont {Cobden},\ and\ \citenamefont {Xu}}]{R12}%
  \BibitemOpen
  \bibfield  {author} {\bibinfo {author} {\bibfnamefont {C.}~\bibnamefont
  {Huang}}, \bibinfo {author} {\bibfnamefont {S.}~\bibnamefont {Wu}}, \bibinfo
  {author} {\bibfnamefont {A.~M.}\ \bibnamefont {Sanchez}}, \bibinfo {author}
  {\bibfnamefont {J.~J.~P.}\ \bibnamefont {Peters}}, \bibinfo {author}
  {\bibfnamefont {R.}~\bibnamefont {Beanland}}, \bibinfo {author}
  {\bibfnamefont {J.~S.}\ \bibnamefont {Ross}}, \bibinfo {author}
  {\bibfnamefont {P.}~\bibnamefont {Rivera}}, \bibinfo {author} {\bibfnamefont
  {W.}~\bibnamefont {Yao}}, \bibinfo {author} {\bibfnamefont {D.~H.}\
  \bibnamefont {Cobden}}, \ and\ \bibinfo {author} {\bibfnamefont
  {X.}~\bibnamefont {Xu}},\ }\href@noop {} {\bibfield  {journal} {\bibinfo
  {journal} {Nat. Mater.}\ }\textbf {\bibinfo {volume} {13}},\ \bibinfo {pages}
  {1096} (\bibinfo {year} {2014})}\BibitemShut {NoStop}%
\bibitem [{\citenamefont {Gong}\ \emph {et~al.}(2014)\citenamefont {Gong},
  \citenamefont {Lin}, \citenamefont {Wang}, \citenamefont {Shi}, \citenamefont
  {Lei}, \citenamefont {Lin}, \citenamefont {Zou}, \citenamefont {G.Ye},
  \citenamefont {Vajtai}, \citenamefont {Yakobsoni}, \citenamefont {Terrones},
  \citenamefont {Terrones}, \citenamefont {Tay}, \citenamefont {Lou},
  \citenamefont {Pantelides}, \citenamefont {Liu}, \citenamefont {Zhou},\ and\
  \citenamefont {Ajayan}}]{R13}%
  \BibitemOpen
  \bibfield  {author} {\bibinfo {author} {\bibfnamefont {Y.}~\bibnamefont
  {Gong}}, \bibinfo {author} {\bibfnamefont {J.}~\bibnamefont {Lin}}, \bibinfo
  {author} {\bibfnamefont {X.}~\bibnamefont {Wang}}, \bibinfo {author}
  {\bibfnamefont {G.}~\bibnamefont {Shi}}, \bibinfo {author} {\bibfnamefont
  {S.}~\bibnamefont {Lei}}, \bibinfo {author} {\bibfnamefont {Z.}~\bibnamefont
  {Lin}}, \bibinfo {author} {\bibfnamefont {X.}~\bibnamefont {Zou}}, \bibinfo
  {author} {\bibnamefont {G.Ye}}, \bibinfo {author} {\bibfnamefont
  {R.}~\bibnamefont {Vajtai}}, \bibinfo {author} {\bibfnamefont {B.~I.}\
  \bibnamefont {Yakobsoni}}, \bibinfo {author} {\bibfnamefont {H.}~\bibnamefont
  {Terrones}}, \bibinfo {author} {\bibfnamefont {M.}~\bibnamefont {Terrones}},
  \bibinfo {author} {\bibfnamefont {B.~K.}\ \bibnamefont {Tay}}, \bibinfo
  {author} {\bibfnamefont {J.}~\bibnamefont {Lou}}, \bibinfo {author}
  {\bibfnamefont {S.~T.}\ \bibnamefont {Pantelides}}, \bibinfo {author}
  {\bibfnamefont {Z.}~\bibnamefont {Liu}}, \bibinfo {author} {\bibfnamefont
  {W.}~\bibnamefont {Zhou}}, \ and\ \bibinfo {author} {\bibfnamefont
  {P.}~\bibnamefont {Ajayan}},\ }\href@noop {} {\bibfield  {journal} {\bibinfo
  {journal} {Nat. Mater.}\ }\textbf {\bibinfo {volume} {13}},\ \bibinfo {pages}
  {1135} (\bibinfo {year} {2014})}\BibitemShut {NoStop}%
\bibitem [{\citenamefont {Zhang}\ \emph
  {et~al.}(2015{\natexlab{a}})\citenamefont {Zhang}, \citenamefont {Lin},
  \citenamefont {Tseng}, \citenamefont {Huang},\ and\ \citenamefont
  {Lee}}]{R14}%
  \BibitemOpen
  \bibfield  {author} {\bibinfo {author} {\bibfnamefont {X.~Q.}\ \bibnamefont
  {Zhang}}, \bibinfo {author} {\bibfnamefont {C.~H.}\ \bibnamefont {Lin}},
  \bibinfo {author} {\bibfnamefont {Y.~W.}\ \bibnamefont {Tseng}}, \bibinfo
  {author} {\bibfnamefont {K.~H.}\ \bibnamefont {Huang}}, \ and\ \bibinfo
  {author} {\bibfnamefont {Y.~H.}\ \bibnamefont {Lee}},\ }\href@noop {}
  {\bibfield  {journal} {\bibinfo  {journal} {Nano Letters}\ }\textbf {\bibinfo
  {volume} {15}},\ \bibinfo {pages} {410} (\bibinfo {year}
  {2015}{\natexlab{a}})}\BibitemShut {NoStop}%
\bibitem [{\citenamefont {Mahjouri-Samani}\ \emph {et~al.}(2015)\citenamefont
  {Mahjouri-Samani}, \citenamefont {Lin}, \citenamefont {Wang}, \citenamefont
  {Lupini}, \citenamefont {Lee}, \citenamefont {Basile}, \citenamefont
  {Boulesbaa}, \citenamefont {Rouleau}, \citenamefont {Puretzky}, \citenamefont
  {Ivanov}, \citenamefont {Xiao}, \citenamefont {Yoon},\ and\ \citenamefont
  {Geohegan}}]{R15}%
  \BibitemOpen
  \bibfield  {author} {\bibinfo {author} {\bibfnamefont {M.}~\bibnamefont
  {Mahjouri-Samani}}, \bibinfo {author} {\bibfnamefont {M.~W.}\ \bibnamefont
  {Lin}}, \bibinfo {author} {\bibfnamefont {K.}~\bibnamefont {Wang}}, \bibinfo
  {author} {\bibfnamefont {A.~R.}\ \bibnamefont {Lupini}}, \bibinfo {author}
  {\bibfnamefont {J.}~\bibnamefont {Lee}}, \bibinfo {author} {\bibfnamefont
  {L.}~\bibnamefont {Basile}}, \bibinfo {author} {\bibfnamefont
  {A.}~\bibnamefont {Boulesbaa}}, \bibinfo {author} {\bibfnamefont {C.~M.}\
  \bibnamefont {Rouleau}}, \bibinfo {author} {\bibfnamefont {A.~A.}\
  \bibnamefont {Puretzky}}, \bibinfo {author} {\bibfnamefont {I.~N.}\
  \bibnamefont {Ivanov}}, \bibinfo {author} {\bibfnamefont {K.}~\bibnamefont
  {Xiao}}, \bibinfo {author} {\bibfnamefont {M.}~\bibnamefont {Yoon}}, \ and\
  \bibinfo {author} {\bibfnamefont {D.~B.}\ \bibnamefont {Geohegan}},\
  }\href@noop {} {\bibfield  {journal} {\bibinfo  {journal} {Nat. Commun.}\
  }\textbf {\bibinfo {volume} {6}},\ \bibinfo {pages} {7749} (\bibinfo {year}
  {2015})}\BibitemShut {NoStop}%
\bibitem [{\citenamefont {Gong}\ \emph {et~al.}(2013)\citenamefont {Gong},
  \citenamefont {Zhang}, \citenamefont {Wang}, \citenamefont {Colombo},
  \citenamefont {Wallace},\ and\ \citenamefont {Cho}}]{R16}%
  \BibitemOpen
  \bibfield  {author} {\bibinfo {author} {\bibfnamefont {C.}~\bibnamefont
  {Gong}}, \bibinfo {author} {\bibfnamefont {H.}~\bibnamefont {Zhang}},
  \bibinfo {author} {\bibfnamefont {W.}~\bibnamefont {Wang}}, \bibinfo {author}
  {\bibfnamefont {L.}~\bibnamefont {Colombo}}, \bibinfo {author} {\bibfnamefont
  {R.~M.}\ \bibnamefont {Wallace}}, \ and\ \bibinfo {author} {\bibfnamefont
  {K.}~\bibnamefont {Cho}},\ }\href@noop {} {\bibfield  {journal} {\bibinfo
  {journal} {Appl. Phys. Lett.}\ }\textbf {\bibinfo {volume} {103}},\ \bibinfo
  {pages} {053513} (\bibinfo {year} {2013})}\BibitemShut {NoStop}%
\bibitem [{\citenamefont {Kosmider}\ and\ \citenamefont {Rossier}(2017)}]{R17}%
  \BibitemOpen
  \bibfield  {author} {\bibinfo {author} {\bibfnamefont {K.}~\bibnamefont
  {Kosmider}}\ and\ \bibinfo {author} {\bibfnamefont {J.~F.}\ \bibnamefont
  {Rossier}},\ }\href@noop {} {\bibfield  {journal} {\bibinfo  {journal} {Phys
  Rev B}\ }\textbf {\bibinfo {volume} {87}},\ \bibinfo {pages} {075451}
  (\bibinfo {year} {2017})}\BibitemShut {NoStop}%
\bibitem [{\citenamefont {Kang}\ \emph {et~al.}(2013)\citenamefont {Kang},
  \citenamefont {Tongay}, \citenamefont {Zhou}, \citenamefont {Li},\ and\
  \citenamefont {Wu}}]{R18}%
  \BibitemOpen
  \bibfield  {author} {\bibinfo {author} {\bibfnamefont {J.}~\bibnamefont
  {Kang}}, \bibinfo {author} {\bibfnamefont {S.}~\bibnamefont {Tongay}},
  \bibinfo {author} {\bibfnamefont {J.}~\bibnamefont {Zhou}}, \bibinfo {author}
  {\bibfnamefont {J.}~\bibnamefont {Li}}, \ and\ \bibinfo {author}
  {\bibfnamefont {J.}~\bibnamefont {Wu}},\ }\href@noop {} {\bibfield  {journal}
  {\bibinfo  {journal} {Appl. Phys. Lett.}\ }\textbf {\bibinfo {volume}
  {102}},\ \bibinfo {pages} {012111} (\bibinfo {year} {2013})}\BibitemShut
  {NoStop}%
\bibitem [{\citenamefont {Ozcelik}\ \emph {et~al.}(2016)\citenamefont
  {Ozcelik}, \citenamefont {Azadani}, \citenamefont {Yang}, \citenamefont
  {Koester},\ and\ \citenamefont {Low}}]{R19}%
  \BibitemOpen
  \bibfield  {author} {\bibinfo {author} {\bibfnamefont {V.~O.}\ \bibnamefont
  {Ozcelik}}, \bibinfo {author} {\bibfnamefont {J.~G.}\ \bibnamefont
  {Azadani}}, \bibinfo {author} {\bibfnamefont {C.}~\bibnamefont {Yang}},
  \bibinfo {author} {\bibfnamefont {S.~J.}\ \bibnamefont {Koester}}, \ and\
  \bibinfo {author} {\bibfnamefont {T.}~\bibnamefont {Low}},\ }\href@noop {}
  {\bibfield  {journal} {\bibinfo  {journal} {Phys. Rev. B}\ }\textbf {\bibinfo
  {volume} {94}},\ \bibinfo {pages} {035125} (\bibinfo {year}
  {2016})}\BibitemShut {NoStop}%
\bibitem [{\citenamefont {Zhang}\ \emph {et~al.}(2016)\citenamefont {Zhang},
  \citenamefont {Xie}, \citenamefont {Peng},\ and\ \citenamefont {Chen}}]{R20}%
  \BibitemOpen
  \bibfield  {author} {\bibinfo {author} {\bibfnamefont {Z.}~\bibnamefont
  {Zhang}}, \bibinfo {author} {\bibfnamefont {Y.}~\bibnamefont {Xie}}, \bibinfo
  {author} {\bibfnamefont {Q.}~\bibnamefont {Peng}}, \ and\ \bibinfo {author}
  {\bibfnamefont {Y.}~\bibnamefont {Chen}},\ }\href@noop {} {\bibfield
  {journal} {\bibinfo  {journal} {Sci. Rep.}\ }\textbf {\bibinfo {volume}
  {6}},\ \bibinfo {pages} {21639} (\bibinfo {year} {2016})}\BibitemShut
  {NoStop}%
\bibitem [{\citenamefont {Avalos-Ovando}, \citenamefont {Mastrogiuseppe},\ and\
  \citenamefont {Ulloa}(2019)}]{R21}%
  \BibitemOpen
  \bibfield  {author} {\bibinfo {author} {\bibfnamefont {O.}~\bibnamefont
  {Avalos-Ovando}}, \bibinfo {author} {\bibfnamefont {D.}~\bibnamefont
  {Mastrogiuseppe}}, \ and\ \bibinfo {author} {\bibfnamefont {S.~E.}\
  \bibnamefont {Ulloa}},\ }\href@noop {} {\bibfield  {journal} {\bibinfo
  {journal} {Phys. Rev. B}\ }\textbf {\bibinfo {volume} {99}},\ \bibinfo
  {pages} {035107} (\bibinfo {year} {2019})}\BibitemShut {NoStop}%
\bibitem [{\citenamefont {Choukroun}\ \emph {et~al.}(2019)\citenamefont
  {Choukroun}, \citenamefont {Pala}, \citenamefont {Fang}, \citenamefont
  {Kaxiras},\ and\ \citenamefont {Dollfus}}]{R22}%
  \BibitemOpen
  \bibfield  {author} {\bibinfo {author} {\bibfnamefont {J.}~\bibnamefont
  {Choukroun}}, \bibinfo {author} {\bibfnamefont {M.}~\bibnamefont {Pala}},
  \bibinfo {author} {\bibfnamefont {S.}~\bibnamefont {Fang}}, \bibinfo {author}
  {\bibfnamefont {E.}~\bibnamefont {Kaxiras}}, \ and\ \bibinfo {author}
  {\bibfnamefont {P.}~\bibnamefont {Dollfus}},\ }\href@noop {} {\bibfield
  {journal} {\bibinfo  {journal} {Nanotechnology}\ }\textbf {\bibinfo {volume}
  {30}},\ \bibinfo {pages} {025201} (\bibinfo {year} {2019})}\BibitemShut
  {NoStop}%
\bibitem [{\citenamefont {Duan}\ \emph
  {et~al.}(2014{\natexlab{b}})\citenamefont {Duan}, \citenamefont {Wang},
  \citenamefont {Shaw}, \citenamefont {Cheng},\ and\ \citenamefont
  {et~al.}}]{R23}%
  \BibitemOpen
  \bibfield  {author} {\bibinfo {author} {\bibfnamefont {X.}~\bibnamefont
  {Duan}}, \bibinfo {author} {\bibfnamefont {C.}~\bibnamefont {Wang}}, \bibinfo
  {author} {\bibfnamefont {J.}~\bibnamefont {Shaw}}, \bibinfo {author}
  {\bibfnamefont {R.}~\bibnamefont {Cheng}}, \ and\ \bibinfo {author}
  {\bibfnamefont {Y.~C.}\ \bibnamefont {et~al.}},\ }\href@noop {} {} (\bibinfo
  {year} {2014}{\natexlab{b}}),\ \Eprint {http://arxiv.org/abs/Nat.
  Nanotechnol.} {Nat. Nanotechnol.} \BibitemShut {NoStop}%
\bibitem [{\citenamefont {Li}\ \emph {et~al.}(2018)\citenamefont {Li},
  \citenamefont {Pu}, \citenamefont {Huang}, \citenamefont {Miyauchi},
  \citenamefont {Matsuda}, \citenamefont {Takenobu},\ and\ \citenamefont
  {Li}}]{R24}%
  \BibitemOpen
  \bibfield  {author} {\bibinfo {author} {\bibfnamefont {M.}~\bibnamefont
  {Li}}, \bibinfo {author} {\bibfnamefont {J.}~\bibnamefont {Pu}}, \bibinfo
  {author} {\bibfnamefont {J.}~\bibnamefont {Huang}}, \bibinfo {author}
  {\bibfnamefont {Y.}~\bibnamefont {Miyauchi}}, \bibinfo {author}
  {\bibfnamefont {K.}~\bibnamefont {Matsuda}}, \bibinfo {author} {\bibfnamefont
  {T.}~\bibnamefont {Takenobu}}, \ and\ \bibinfo {author} {\bibfnamefont
  {L.}~\bibnamefont {Li}},\ }\href@noop {} {\bibfield  {journal} {\bibinfo
  {journal} {Adv. Funct. Mater.}\ }\textbf {\bibinfo {volume} {28}},\ \bibinfo
  {pages} {1706860} (\bibinfo {year} {2018})}\BibitemShut {NoStop}%
\bibitem [{\citenamefont {Chen}\ \emph {et~al.}(2019)\citenamefont {Chen},
  \citenamefont {Hofmann}, \citenamefont {Yao}, \citenamefont {Chiu},
  \citenamefont {Chen}, \citenamefont {Luo}, \citenamefont {Hsu},\ and\
  \citenamefont {Hsieh}}]{R25}%
  \BibitemOpen
  \bibfield  {author} {\bibinfo {author} {\bibfnamefont {D.}~\bibnamefont
  {Chen}}, \bibinfo {author} {\bibfnamefont {M.}~\bibnamefont {Hofmann}},
  \bibinfo {author} {\bibfnamefont {H.}~\bibnamefont {Yao}}, \bibinfo {author}
  {\bibfnamefont {S.}~\bibnamefont {Chiu}}, \bibinfo {author} {\bibfnamefont
  {S.}~\bibnamefont {Chen}}, \bibinfo {author} {\bibfnamefont {Y.}~\bibnamefont
  {Luo}}, \bibinfo {author} {\bibfnamefont {C.}~\bibnamefont {Hsu}}, \ and\
  \bibinfo {author} {\bibfnamefont {Y.}~\bibnamefont {Hsieh}},\ }\href@noop {}
  {\bibfield  {journal} {\bibinfo  {journal} {ACS Appl. Mater. Interfaces}\
  }\textbf {\bibinfo {volume} {11}},\ \bibinfo {pages} {6384} (\bibinfo {year}
  {2019})}\BibitemShut {NoStop}%
\bibitem [{\citenamefont {Amani}\ \emph {et~al.}(2015)\citenamefont {Amani},
  \citenamefont {Lien}, \citenamefont {Kiriya}, \citenamefont {Xiao},
  \citenamefont {Azcatl}, \citenamefont {Noh}, \citenamefont {Madhvapathy},
  \citenamefont {Addou}, \citenamefont {Santosh},\ and\ \citenamefont
  {Dubey}}]{R26}%
  \BibitemOpen
  \bibfield  {author} {\bibinfo {author} {\bibfnamefont {M.}~\bibnamefont
  {Amani}}, \bibinfo {author} {\bibfnamefont {D.~H.}\ \bibnamefont {Lien}},
  \bibinfo {author} {\bibfnamefont {D.}~\bibnamefont {Kiriya}}, \bibinfo
  {author} {\bibfnamefont {J.}~\bibnamefont {Xiao}}, \bibinfo {author}
  {\bibfnamefont {A.}~\bibnamefont {Azcatl}}, \bibinfo {author} {\bibfnamefont
  {J.}~\bibnamefont {Noh}}, \bibinfo {author} {\bibfnamefont {S.~R.}\
  \bibnamefont {Madhvapathy}}, \bibinfo {author} {\bibfnamefont
  {R.}~\bibnamefont {Addou}}, \bibinfo {author} {\bibfnamefont
  {K.}~\bibnamefont {Santosh}}, \ and\ \bibinfo {author} {\bibfnamefont
  {M.}~\bibnamefont {Dubey}},\ }\href@noop {} {\bibfield  {journal} {\bibinfo
  {journal} {Science}\ }\textbf {\bibinfo {volume} {4}},\ \bibinfo {pages}
  {350} (\bibinfo {year} {2015})}\BibitemShut {NoStop}%
\bibitem [{\citenamefont {Godde}\ \emph {et~al.}(2016)\citenamefont {Godde},
  \citenamefont {Schmidt}, \citenamefont {Schmutzler}, \citenamefont {Asmann},
  \citenamefont {Debus}, \citenamefont {Withers}, \citenamefont {Alexeev},
  \citenamefont {Pozo-Zamudio}, \citenamefont {Skrypka}, \citenamefont
  {Novoselov}, \citenamefont {Bayer},\ and\ \citenamefont
  {Tartakovskii}}]{R27}%
  \BibitemOpen
  \bibfield  {author} {\bibinfo {author} {\bibfnamefont {T.}~\bibnamefont
  {Godde}}, \bibinfo {author} {\bibfnamefont {D.}~\bibnamefont {Schmidt}},
  \bibinfo {author} {\bibfnamefont {J.}~\bibnamefont {Schmutzler}}, \bibinfo
  {author} {\bibfnamefont {M.}~\bibnamefont {Asmann}}, \bibinfo {author}
  {\bibfnamefont {J.}~\bibnamefont {Debus}}, \bibinfo {author} {\bibfnamefont
  {F.}~\bibnamefont {Withers}}, \bibinfo {author} {\bibfnamefont {E.~M.}\
  \bibnamefont {Alexeev}}, \bibinfo {author} {\bibfnamefont {O.~D.}\
  \bibnamefont {Pozo-Zamudio}}, \bibinfo {author} {\bibfnamefont {O.~V.}\
  \bibnamefont {Skrypka}}, \bibinfo {author} {\bibfnamefont {K.~S.}\
  \bibnamefont {Novoselov}}, \bibinfo {author} {\bibfnamefont {M.}~\bibnamefont
  {Bayer}}, \ and\ \bibinfo {author} {\bibfnamefont {A.~I.}\ \bibnamefont
  {Tartakovskii}},\ }\href@noop {} {\bibfield  {journal} {\bibinfo  {journal}
  {Phys. Rev. B}\ }\textbf {\bibinfo {volume} {94}},\ \bibinfo {pages} {165301}
  (\bibinfo {year} {2016})}\BibitemShut {NoStop}%
\bibitem [{\citenamefont {Zhang}\ \emph
  {et~al.}(2015{\natexlab{b}})\citenamefont {Zhang}, \citenamefont {You},
  \citenamefont {Zhao},\ and\ \citenamefont {Heinz}}]{R28}%
  \BibitemOpen
  \bibfield  {author} {\bibinfo {author} {\bibfnamefont {X.}~\bibnamefont
  {Zhang}}, \bibinfo {author} {\bibfnamefont {Y.}~\bibnamefont {You}}, \bibinfo
  {author} {\bibfnamefont {S.~Y.~F.}\ \bibnamefont {Zhao}}, \ and\ \bibinfo
  {author} {\bibfnamefont {T.~F.}\ \bibnamefont {Heinz}},\ }\href@noop {}
  {\bibfield  {journal} {\bibinfo  {journal} {Phys. Rev. Lett.}\ }\textbf
  {\bibinfo {volume} {115}},\ \bibinfo {pages} {257403} (\bibinfo {year}
  {2015}{\natexlab{b}})}\BibitemShut {NoStop}%
\bibitem [{\citenamefont {Rivera}, \citenamefont {Yao},\ and\ \citenamefont
  {et~al.}(2015)}]{R29}%
  \BibitemOpen
  \bibfield  {author} {\bibinfo {author} {\bibfnamefont {P.}~\bibnamefont
  {Rivera}}, \bibinfo {author} {\bibfnamefont {W.}~\bibnamefont {Yao}}, \ and\
  \bibinfo {author} {\bibfnamefont {X.~D.~X.}\ \bibnamefont {et~al.}},\
  }\href@noop {} {\bibfield  {journal} {\bibinfo  {journal} {., Nat. Commun.}\
  }\textbf {\bibinfo {volume} {6}},\ \bibinfo {pages} {6242} (\bibinfo {year}
  {2015})}\BibitemShut {NoStop}%
\bibitem [{\citenamefont {Li}\ \emph {et~al.}(2015)\citenamefont {Li},
  \citenamefont {Li}, \citenamefont {Shi}, \citenamefont {Zhang}, \citenamefont
  {Yang},\ and\ \citenamefont {Chang}}]{R30}%
  \BibitemOpen
  \bibfield  {author} {\bibinfo {author} {\bibfnamefont {Y.~M.}\ \bibnamefont
  {Li}}, \bibinfo {author} {\bibfnamefont {J.}~\bibnamefont {Li}}, \bibinfo
  {author} {\bibfnamefont {L.~K.}\ \bibnamefont {Shi}}, \bibinfo {author}
  {\bibfnamefont {D.}~\bibnamefont {Zhang}}, \bibinfo {author} {\bibfnamefont
  {W.}~\bibnamefont {Yang}}, \ and\ \bibinfo {author} {\bibfnamefont
  {K.}~\bibnamefont {Chang}},\ }\href@noop {} {\bibfield  {journal} {\bibinfo
  {journal} {Phys. Rev. Lett.}\ }\textbf {\bibinfo {volume} {115}},\ \bibinfo
  {pages} {166804} (\bibinfo {year} {2015})}\BibitemShut {NoStop}%
\bibitem [{\citenamefont {Latini}, \citenamefont {Olsen},\ and\ \citenamefont
  {Thygesen}(2015)}]{R31}%
  \BibitemOpen
  \bibfield  {author} {\bibinfo {author} {\bibfnamefont {S.}~\bibnamefont
  {Latini}}, \bibinfo {author} {\bibfnamefont {T.}~\bibnamefont {Olsen}}, \
  and\ \bibinfo {author} {\bibfnamefont {K.~S.}\ \bibnamefont {Thygesen}},\
  }\href@noop {} {\bibfield  {journal} {\bibinfo  {journal} {Phys. Rev. B}\
  }\textbf {\bibinfo {volume} {92}},\ \bibinfo {pages} {245123} (\bibinfo
  {year} {2015})}\BibitemShut {NoStop}%
\bibitem [{\citenamefont {Lau}\ \emph {et~al.}(2018)\citenamefont {Lau},
  \citenamefont {Calvin}, \citenamefont {Gong}, \citenamefont {Yu},\ and\
  \citenamefont {Yao}}]{R33}%
  \BibitemOpen
  \bibfield  {author} {\bibinfo {author} {\bibfnamefont {K.~W.}\ \bibnamefont
  {Lau}}, \bibinfo {author} {\bibnamefont {Calvin}}, \bibinfo {author}
  {\bibfnamefont {Z.}~\bibnamefont {Gong}}, \bibinfo {author} {\bibfnamefont
  {H.}~\bibnamefont {Yu}}, \ and\ \bibinfo {author} {\bibfnamefont
  {W.}~\bibnamefont {Yao}},\ }\href@noop {} {\bibfield  {journal} {\bibinfo
  {journal} {Phys. Rev. B}\ }\textbf {\bibinfo {volume} {98}},\ \bibinfo
  {pages} {115427} (\bibinfo {year} {2018})}\BibitemShut {NoStop}%
\bibitem [{\citenamefont {Jin}\ \emph {et~al.}(2019)\citenamefont {Jin},
  \citenamefont {Michaud}, \citenamefont {Gong}, \citenamefont {Wan},
  \citenamefont {Wei},\ and\ \citenamefont {Guo}}]{R32}%
  \BibitemOpen
  \bibfield  {author} {\bibinfo {author} {\bibfnamefont {H.}~\bibnamefont
  {Jin}}, \bibinfo {author} {\bibfnamefont {V.}~\bibnamefont {Michaud}},
  \bibinfo {author} {\bibfnamefont {Z.~R.}\ \bibnamefont {Gong}}, \bibinfo
  {author} {\bibfnamefont {L.}~\bibnamefont {Wan}}, \bibinfo {author}
  {\bibfnamefont {Y.}~\bibnamefont {Wei}}, \ and\ \bibinfo {author}
  {\bibfnamefont {H.}~\bibnamefont {Guo}},\ }\href@noop {} {\bibfield
  {journal} {\bibinfo  {journal} {J. Mater. Chemi.}\ }\textbf {\bibinfo
  {volume} {7}},\ \bibinfo {pages} {13} (\bibinfo {year} {2019})}\BibitemShut
  {NoStop}%
\bibitem [{\citenamefont {Sasaki}, \citenamefont {Murakami},\ and\
  \citenamefont {Saito}(2006)}]{R34}%
  \BibitemOpen
  \bibfield  {author} {\bibinfo {author} {\bibfnamefont {K.}~\bibnamefont
  {Sasaki}}, \bibinfo {author} {\bibfnamefont {S.}~\bibnamefont {Murakami}}, \
  and\ \bibinfo {author} {\bibfnamefont {R.}~\bibnamefont {Saito}},\
  }\href@noop {} {\bibfield  {journal} {\bibinfo  {journal} {J. Phys. Soc.
  Jpn}\ }\textbf {\bibinfo {volume} {75}},\ \bibinfo {pages} {7} (\bibinfo
  {year} {2006})}\BibitemShut {NoStop}%
\bibitem [{\citenamefont {Liu}\ \emph {et~al.}(2014)\citenamefont {Liu},
  \citenamefont {Shan}, \citenamefont {Yao}, \citenamefont {Yao},\ and\
  \citenamefont {Xiao}}]{R35}%
  \BibitemOpen
  \bibfield  {author} {\bibinfo {author} {\bibfnamefont {G.~B.}\ \bibnamefont
  {Liu}}, \bibinfo {author} {\bibfnamefont {W.~Y.}\ \bibnamefont {Shan}},
  \bibinfo {author} {\bibfnamefont {Y.}~\bibnamefont {Yao}}, \bibinfo {author}
  {\bibfnamefont {W.}~\bibnamefont {Yao}}, \ and\ \bibinfo {author}
  {\bibfnamefont {D.}~\bibnamefont {Xiao}},\ }\href@noop {} {\bibfield
  {journal} {\bibinfo  {journal} {Phys. Rev. B}\ }\textbf {\bibinfo {volume}
  {89}},\ \bibinfo {pages} {039901} (\bibinfo {year} {2014})}\BibitemShut
  {NoStop}%
\bibitem [{\citenamefont {Cudazzo}, \citenamefont {Tokatly},\ and\
  \citenamefont {Rubio}(2011)}]{R36}%
  \BibitemOpen
  \bibfield  {author} {\bibinfo {author} {\bibfnamefont {P.}~\bibnamefont
  {Cudazzo}}, \bibinfo {author} {\bibfnamefont {I.~V.}\ \bibnamefont
  {Tokatly}}, \ and\ \bibinfo {author} {\bibfnamefont {A.}~\bibnamefont
  {Rubio}},\ }\href@noop {} {\bibfield  {journal} {\bibinfo  {journal} {Phys.
  Rev. B}\ }\textbf {\bibinfo {volume} {84}},\ \bibinfo {pages} {085406}
  (\bibinfo {year} {2011})}\BibitemShut {NoStop}%
\bibitem [{\citenamefont {Hai}(2011)}]{R37}%
  \BibitemOpen
  \bibfield  {author} {\bibinfo {author} {\bibfnamefont {D.~S.}\ \bibnamefont
  {Hai}},\ }\href@noop {} {\bibfield  {journal} {\bibinfo  {journal} {Commum.
  Theor. Phys.}\ }\textbf {\bibinfo {volume} {55}},\ \bibinfo {pages} {969}
  (\bibinfo {year} {2011})}\BibitemShut {NoStop}%
\bibitem [{\citenamefont {Sous}\ and\ \citenamefont {Alhaidari}(2016)}]{R38}%
  \BibitemOpen
  \bibfield  {author} {\bibinfo {author} {\bibfnamefont {A.~J.}\ \bibnamefont
  {Sous}}\ and\ \bibinfo {author} {\bibfnamefont {A.~D.}\ \bibnamefont
  {Alhaidari}},\ }\href@noop {} {\bibfield  {journal} {\bibinfo  {journal} {J.
  Appl. Mathe. and Phys.}\ }\textbf {\bibinfo {volume} {4}},\ \bibinfo {pages}
  {79} (\bibinfo {year} {2016})}\BibitemShut {NoStop}%
\bibitem [{\citenamefont {Gonzalez-Espimoza}\ and\ \citenamefont
  {Ayers}(2016)}]{R39}%
  \BibitemOpen
  \bibfield  {author} {\bibinfo {author} {\bibfnamefont {C.~E.}\ \bibnamefont
  {Gonzalez-Espimoza}}\ and\ \bibinfo {author} {\bibfnamefont {P.~W.}\
  \bibnamefont {Ayers}},\ }\href@noop {} {\bibfield  {journal} {\bibinfo
  {journal} {Theor. Chemist. Accou.}\ }\textbf {\bibinfo {volume} {135}},\
  \bibinfo {pages} {256} (\bibinfo {year} {2016})}\BibitemShut {NoStop}%
\bibitem [{\citenamefont {Sous}\ and\ \citenamefont {EI-Kawni}(2018)}]{R40}%
  \BibitemOpen
  \bibfield  {author} {\bibinfo {author} {\bibfnamefont {A.~J.}\ \bibnamefont
  {Sous}}\ and\ \bibinfo {author} {\bibfnamefont {M.~I.}\ \bibnamefont
  {EI-Kawni}},\ }\href@noop {} {\bibfield  {journal} {\bibinfo  {journal} {J.
  Appl. Mathe. and Phys.}\ }\textbf {\bibinfo {volume} {6}},\ \bibinfo {pages}
  {901} (\bibinfo {year} {2018})}\BibitemShut {NoStop}%
\bibitem [{\citenamefont {Chaves}\ and\ \citenamefont {Jimenez}(2018)}]{R41}%
  \BibitemOpen
  \bibfield  {author} {\bibinfo {author} {\bibfnamefont {F.~A.}\ \bibnamefont
  {Chaves}}\ and\ \bibinfo {author} {\bibfnamefont {D.}~\bibnamefont
  {Jimenez}},\ }\href@noop {} {\bibfield  {journal} {\bibinfo  {journal}
  {Nanotech.}\ }\textbf {\bibinfo {volume} {29}},\ \bibinfo {pages} {27}
  (\bibinfo {year} {2018})}\BibitemShut {NoStop}%
\bibitem [{\citenamefont {Felbacq}\ and\ \citenamefont {Rousseau}(2019)}]{R42}%
  \BibitemOpen
  \bibfield  {author} {\bibinfo {author} {\bibfnamefont {D.}~\bibnamefont
  {Felbacq}}\ and\ \bibinfo {author} {\bibfnamefont {E.}~\bibnamefont
  {Rousseau}},\ }\href@noop {} {\bibfield  {journal} {\bibinfo  {journal} {Ann.
  Phys. (Berlin)}\ }\textbf {\bibinfo {volume} {531}},\ \bibinfo {pages}
  {1800486} (\bibinfo {year} {2019})}\BibitemShut {NoStop}%
\bibitem [{\citenamefont {Zhang}\ and\ \citenamefont {Ma}(2019)}]{R43}%
  \BibitemOpen
  \bibfield  {author} {\bibinfo {author} {\bibfnamefont {J.~Z.}\ \bibnamefont
  {Zhang}}\ and\ \bibinfo {author} {\bibfnamefont {J.~Z.}\ \bibnamefont {Ma}},\
  }\href@noop {} {\bibfield  {journal} {\bibinfo  {journal} {Conden. Matt.}\
  }\textbf {\bibinfo {volume} {31}},\ \bibinfo {pages} {10} (\bibinfo {year}
  {2019})}\BibitemShut {NoStop}%
\bibitem [{\citenamefont {Pulci}\ \emph {et~al.}(2015)\citenamefont {Pulci},
  \citenamefont {Marsili}, \citenamefont {Garbuio}, \citenamefont {Gori},
  \citenamefont {Kupchak},\ and\ \citenamefont {Bechstedt}}]{R44}%
  \BibitemOpen
  \bibfield  {author} {\bibinfo {author} {\bibfnamefont {O.}~\bibnamefont
  {Pulci}}, \bibinfo {author} {\bibfnamefont {M.}~\bibnamefont {Marsili}},
  \bibinfo {author} {\bibfnamefont {V.}~\bibnamefont {Garbuio}}, \bibinfo
  {author} {\bibfnamefont {P.}~\bibnamefont {Gori}}, \bibinfo {author}
  {\bibfnamefont {I.}~\bibnamefont {Kupchak}}, \ and\ \bibinfo {author}
  {\bibfnamefont {F.}~\bibnamefont {Bechstedt}},\ }\href@noop {} {\bibfield
  {journal} {\bibinfo  {journal} {Phys. Stat. Sol. B}\ }\textbf {\bibinfo
  {volume} {252}},\ \bibinfo {pages} {72} (\bibinfo {year} {2015})}\BibitemShut
  {NoStop}%
\bibitem [{\citenamefont {Xiao}\ \emph {et~al.}(2012)\citenamefont {Xiao},
  \citenamefont {Liu}, \citenamefont {Feng}, \citenamefont {Xu},\ and\
  \citenamefont {Yao}}]{R46}%
  \BibitemOpen
  \bibfield  {author} {\bibinfo {author} {\bibfnamefont {D.}~\bibnamefont
  {Xiao}}, \bibinfo {author} {\bibfnamefont {G.~B.}\ \bibnamefont {Liu}},
  \bibinfo {author} {\bibfnamefont {W.}~\bibnamefont {Feng}}, \bibinfo {author}
  {\bibfnamefont {X.}~\bibnamefont {Xu}}, \ and\ \bibinfo {author}
  {\bibfnamefont {W.}~\bibnamefont {Yao}},\ }\href@noop {} {\bibfield
  {journal} {\bibinfo  {journal} {Phys. Rev. Lett.}\ }\textbf {\bibinfo
  {volume} {108}},\ \bibinfo {pages} {196802} (\bibinfo {year}
  {2012})}\BibitemShut {NoStop}%
\bibitem [{\citenamefont {Evans}(2007)}]{R47}%
  \BibitemOpen
  \bibfield  {author} {\bibinfo {author} {\bibfnamefont {D.~J.}\ \bibnamefont
  {Evans}},\ }\href@noop {} {\bibfield  {journal} {\bibinfo  {journal} {Int. J.
  Comput. Math.}\ }\textbf {\bibinfo {volume} {39}},\ \bibinfo {pages} {1}
  (\bibinfo {year} {2007})}\BibitemShut {NoStop}%
\end{thebibliography}%

\end{document}